

\input harvmac
\input epsf

\noblackbox
\def\epm#1#2{\hbox{${+#1}\atop {-#2}$}}
\pageno=0\nopagenumbers\tolerance=10000\hfuzz=5pt
\line{\hfill CERN-TH/95-148}
\line{\hfill \tt hep-ph/9506233}
\vskip 24pt
\centerline{\bf  DETERMINATION OF $\alpha_s$ FROM $F_2^p$ AT HERA}
\vskip 36pt\centerline{Richard~D.~Ball\footnote{*}{On leave
 from a Royal Society University Research Fellowship.}
 and Stefano~Forte\footnote{\dag}{On leave
 from INFN, Sezione di Torino, Italy.}}
\vskip 12pt
\centerline{\it Theory Division, CERN,}
\centerline{\it CH-1211 Gen\`eve 23, Switzerland.}
\vskip 36pt
{\centerline{\bf Abstract}
\medskip\narrower
\ninepoint\baselineskip=9pt plus 2pt minus 1pt
\lineskiplimit=1pt \lineskip=2pt
We compute the proton structure function $F_2^p$ at small $x$ and
large $Q^2$ at next-to-leading order in $\alpha_s(Q^2)$, including
summations of all leading and subleading logarithms of $Q^2$ and
$1/x$ in a way consistent with  momentum conservation. We perform
a detailed comparison to the 1993 HERA data, and show that they may
be used to determine $\alpha_s(M_Z^2)=0.120\pm 0.005 {\rm (exp)}
\pm{0.009} {\rm (th)}$. The theoretical error is dominated by the
renormalization and factorization scheme ambiguities.
\smallskip}
\vskip 20pt \centerline{Submitted to: {\it Physics Letters B}}
\vskip 24pt
\line{CERN-TH/95-148\hfill}
\line{\tt hep-ph/9506233\hfill}
\line{May 1995\hfill}

\vfill\eject \footline={\hss\tenrm\folio\hss}


\def\frac#1#2{{{#1}\over {#2}}}
\def\half{\hbox{${1\over 2}$}}
\def\quarter{\hbox{${1\over 4}$}}
\def\smallfrac#1#2{\hbox{${{#1}\over {#2}}$}}

\def\MeV{{\rm MeV}}\def\GeV{{\rm GeV}}

\def\DZP{\hbox{D0$'$}}\def\DMP{\hbox{D-$'$}}
\def\MS{\hbox{$\overline{\rm MS}$}}
\def\QMS{Q$_0$\MS}
\def\QDIS{Q$_0$DIS}
\catcode`@=11 
\def\slash#1{\mathord{\mathpalette\c@ncel#1}}
 \def\c@ncel#1#2{\ooalign{$\hfil#1\mkern1mu/\hfil$\crcr$#1#2$}}
\def\lsim{\mathrel{\mathpalette\@versim<}}
\def\gsim{\mathrel{\mathpalette\@versim>}}
 \def\@versim#1#2{\lower0.2ex\vbox{\baselineskip\z@skip\lineskip\z@skip
       \lineskiplimit\z@\ialign{$\m@th#1\hfil##$\crcr#2\crcr\sim\crcr}}}
\catcode`@=12 

\def\coeff{\gamma}
\def\PR{{\it Phys.~Rev.~}}

\def\NP{{\it Nucl.~Phys.~}}
\def\NPBPS{{\it Nucl.~Phys.~B (Proc.~Suppl.)~}}
\def\PL{{\it Phys.~Lett.~}}

\def\ZP{{\it Zeit.~Phys.~}}

\def\vol#1{{\bf #1}}\def\vyp#1#2#3{\vol{#1} (#2) #3}


\nref\ZEUS{ZEUS Collaboration, \ZP\vyp{C65}{1995}{379}.}
\nref\Hone{H1 Collaboration, \NP\vyp{B439}{1995}{471}.}
Experimental knowledge of the proton structure function $F_2$ at
large $Q^2$ and small $x$ has dramatically improved since data from
HERA
have become available~\refs{\ZEUS,\Hone}. Whereas the data are in beautiful
agreement~\ref\Test{R.~D.~Ball and S.~Forte, \PL\vyp{B336}{1994}{77}.}
with the leading-order prediction of perturbative
QCD~\ref\DGPTWZ{A.~De~Rujula, S.L.~Glashow, H.D.~Politzer,
S.B.~Treiman, F.~Wilczek and A.~Zee, \PR\vyp{D10}{1974}{1649}.}, and in
particular display the double scaling behaviour which
follows~\ref\DAS{R.~D.~Ball and S.~Forte, \PL\vyp{B335}{1994}{77}.}
 from it, the improvement in experimental accuracy now opens up
the possibility of more detailed tests of the theory. Indeed, reliable
perturbative calculations
in the small $x$ region turn out to be possible thanks to the fact that
perturbation theory can be reorganized to give an evolution equation
which provides  order by order a convergent sum of all leading logarithms of
$x$ and
$Q^2$~\ref\Summing{R.D.~Ball and S.~Forte,
{\it Phys. Lett.} {\bf B351} (1995) 313.}.
Here we will show that full next-to leading
order (NLO) computations give an excellent description of all the
HERA data. Furthermore, this description is largely independent
of any detailed nonperturbative (thus uncalculable) information, such
as the detailed shape of input parton distributions, but is relatively
sensitive to the precise value of the strong coupling $\alpha_s$,
which may thus be determined using presently available
data~\ref\Moriond{R.D.~Ball and S.~Forte, CERN-TH/95-132 {\tt
hep-ph/9505388} to be published in the proceedings of the
XXXth Rencontres de Moriond, Les Arcs, March 1995.}.

The determination of $\alpha_s$ from the shape of $F_2^p$ at small $x$
which we will discuss here is similar to that used in the
analysis of large $x$ structure functions, however it also shares some of
the features of measurements based on less inclusive quantities
(see ref.~\ref\alfrev{
G.~Altarelli in `QCD - 20 Years Later', Aachen, 1992\semi
S.~Catani in `International Europhysics Conference on High Energy
Physics', Marseille, 1993\semi
S.~Bethke in `QCD94', Montpellier, 1994\semi
B.R.~Webber in `XXVII International Conference on High Energy
Physics', Glasgow, 1994.}).
In fact, many of the more inclusive measurements (for example $e^+e^- \to$
hadrons)
require precise data with very high statistics, since it is necessary
to measure the small $Q^2$-dependent deviations from an
$\alpha_s$-independent leading order result. The determinations from
scaling violations in structure functions obtained from fixed target
deep inelastic scattering are even more difficult, since there the
leading behaviour (the parton distribution functions at some initial scale)
cannot be reliably computed, but must be fitted.
 By contrast,
measurements based on less inclusive quantities which are essentially
proportional to $\alpha_s$ at leading order (for example jet
rates, event shapes and energy correlations) can
be made with much lower statistics: the dominant error then comes
 from hadronization uncertainties.
In this respect the determination of $\alpha_s$ from $F_2^p$ at small $x$
discussed here is akin to the
jet determination, and, in particular, it can be made
with relatively low statistics;
however here there are no hadronization effects, and higher twist
corrections will turn out to be relatively small.

The possibility of determining $\alpha_s$ from the presently available
HERA data \refs{\ZEUS,\Hone} for $F_2^p(x,Q^2)$ follows directly from the
empirical fact that the data exhibit double scaling \DAS:
throughout the HERA kinematic region $\ln F_2^p$ rises linearly in
$\sigma\equiv\sqrt{\xi\zeta}$, but is independent of
$\rho\equiv\sqrt{\xi/\zeta}$ (where, as in ref.\DAS,
$\xi\equiv\ln\smallfrac{x_0}{x}$,
$\zeta\equiv\ln\smallfrac{t}{t_0}$, $t\equiv\ln Q^2/\Lambda^2$).
More precisely, $R_F F_2$, where
$R_F$ is proportional to $\sigma^{1/2}\rho\exp(-2\gamma\sigma+
\delta\sigma/\rho)$, is independent of both $\sigma$ and $\rho$. This
may be seen immediately from the scaling plots \fig\scaling{Double
scaling plots of the 1993 HERA data on $F_2^p$: i)
$R_F F_2^p$ vs. $\sigma$ for a) the ZEUS data \ZEUS, b) the H1 data
\Hone; ii) the same vs. $\rho$. The data plotted are those included
in the fits but with the further cut
$Q^2 < 100 \GeV^2$; $R_F$ is accordingly
evaluated with $n_f=4$ and (as in \DAS) $Q_0=1\GeV$, $x_0=0.1$ and
$\Lambda = 260\MeV$. The dotted, dashed, dotdash, and solid curves correspond
respectively to double scaling, two loops, double leading \QDIS, and
double leading DIS, with $Q_0=2\GeV$, $x_0=0.1$, and the remaining
parameters fitted as in table~6. The curves in the $\sigma$-plots
have $\rho=2.2$: those in the $\rho$-plots $\sigma=1.7$.}. The two parameters
$\gamma\equiv \sqrt{12/\beta_0}$, $\delta\equiv
(11+\smallfrac{2n_f}{27})\big/\beta_0$ depend inversely on the rate
of running of $\alpha_s(t)=\smallfrac{4\pi}{\beta_0
t}+O(\smallfrac{1}{t^2})$; $\beta_0=11-\smallfrac{2}{3}n_f$ is the
leading coefficient of the $\beta$-function. The fact that the HERA data
exhibit double scaling is thus direct evidence for the
running of $\alpha_s(t)$ \Test. Furthermore,
the  growth of $F_2$ at small $x$ and large $Q^2$ is due
primarily to the collinear singularity in the triple gluon
vertex \DGPTWZ\ (at leading order in perturbation theory), and it is thus
directly and simply related to the actual value of $\alpha_s$. This is
seen most clearly by rewriting the predicted~\refs{\DGPTWZ,\DAS}
asymptotic leading order
behaviour of $F_2^p$ as
\eqn\Ftwoas
{F_2^p\sim {\cal N}
\xi^{-3/4}\Big[\ln\frac{\alpha_s(t_0)}{\alpha_s(t)}\Big]^{1/4}
\Big(\frac{\alpha_s(t)}{\alpha_s(t_0)}\Big)^{\delta}
\exp\Bigg\{2\gamma\sqrt{\xi}\sqrt{\ln\frac{\alpha_s(t_0)}{\alpha_s(t)}}\Bigg\}
\Big(1+O(\alpha_s(t))\Big).}
Since the third term rises more slowly than any power of $\alpha_s$,
at very large $t$ (and fixed $x$) the second term dominates and Bjorken
scaling is eventually recovered. However at smaller values of $t$ and
small enough values of $x$ (and in particularly in the HERA region
\DAS) the strong growth in the third term overwhelms the power
fall-off in the second, and $F_2$ rises quickly as $\alpha_s$ falls.
It is this strong dependence of the leading behaviour of $F_2$ at
small $x$ on the value of $\alpha_s$ which makes it possible to make a precise
determination of $\alpha_s$ using the existing HERA data~\refs{\ZEUS,\Hone}.
Indeed, fitting the functional form \Ftwoas\ to the data, fixing $Q_0=1~\GeV$
and $x_0=0.1$, as in \DAS, gives $\alpha_s(M_Z)=0.125\pm 0.002$ (at
one loop), with a $\chi^2$ of $68/121$, which is at least encouraging.

Of course, to actually determine $\alpha_s$ requires at least a full
NLO computation, in order to fix the scale.
Indeed, a two loop computation~\ref\Mont{R.D.~Ball and S.~Forte,
   in the proceedings of {\it ``QCD94''}, Montpellier, July 1994
   (\NPBPS\vyp{39B,C}{1995}{25}).} reveals that NLO effects lead to
a reduction in rate of growth of $F_2$ in the HERA region: although
the asymptotic slope $2\gamma$ is approached at very large $\sigma$,
subasymptotically  the slope is somewhat lower.
This is essentially due to the fact that the leading singularity in
the two-loop splitting function $P^{gg}_1$ is of opposite sign to that
at leading order: in \MS\ scheme
\eqn\pggsing{
P^{gg}(x,t) = \Big[2C_A - \frac{(46C_A-12C_F)T_R n_f}{9}
\frac{\alpha_s(t)}{2\pi}\Big]\frac{1}{x} + \cdots,}
where $C_A = 3$, $C_F = \smallfrac{4}{3}$, $T_R = \half$ for $N_c=3$.
The slope of the rise in $F_2$ is accordingly reduced. Indeed
this reduction can now be seen in the data (see table~4 and \scaling):
agreement with the data is significantly improved, essentially because
of the progressive fall in the $\sigma$ plot. This suggests that a
consistent (i.e. reasonably scale independent) determination of $\alpha_s$ is
indeed going to be possible.


However, in the small $x$ region, i.e. when two large scales
are present ($Q^2$ and  the centre of mass
energy of the partonic scattering process $s={1-x\over x}Q^2$)
a NLO computation may be performed
in a variety of inequivalent expansion schemes, which correspond to
different ways of organizing the perturbative expansion~\Summing.
The standard two loop computation, where only logs of $Q^2$ are considered
leading, is one such expansion scheme, whereas other schemes are
obtained by also
including logs of $1/x$. Unlike different factorization or
renormalization schemes, which are equivalent up to  higher order
perturbative corrections, different expansion schemes can lead
to physically inequivalent predictions because they sum up different
classes of logarithms. A comparison with the data may thus allow one
to decide in which region each expansion scheme is physically
most appropriate.

The fact that the HERA data agree well with
the double scaling prediction \Ftwoas, which treats the two large scales
on the same footing (retaining only terms with the same number of powers of
$\log Q^2$, $\log {1\over x}$, and $\alpha_s$)
suggests that the most appropriate expansion scheme
in the HERA region
is the double leading scheme~\Summing, where one sums all contributions
to splitting functions where each power of $\alpha_s$ is accompanied by
either a power of $\log Q^2$ or  $\log {1\over x}$, i.e. the two scales
are still treated symmetrically.
At leading order this means that one should keep all terms with either one
power of $\alpha_s$ or any number of powers of
$\alpha_s\log\smallfrac{1}{x}$, and at NLO all
terms with an extra power of $\alpha_s$. The singlet splitting
functions are then of the form\foot{In the standard notation in which
the evolution of the singlet parton distribution functions $f^i(x,t)$
is given by $\frac{d}{dt}f^i = \frac{\alpha_s}{2\pi}P^{ij}\otimes f^j$:
different conventions were used in refs.\refs{\DAS,\Summing}.}
\eqn\split{\eqalign{
&P^{ij}(x,t) = P^{ij}_{\rm LO}(x)+\smallfrac{\alpha_s(t)}{2\pi}P^{ij}_{\rm NLO}
(x,t)+\cdots,\cr
&P^{ij}_{\rm LO}(x)=P^{ij}_1(x)+P^{ij}_s(x,t)
\qquad P^{ij}_{\rm NLO}=P^{ij}_2(x)+P^{ij}_{ss}(x,t),\cr}}
where $P^{ij}_1(x)$ and $P^{ij}_2(x)$ are the usual one and two loop
splitting functions, while $P^{ij}_s(x,t)$ and $P^{ij}_{ss}(x,t)$ are
(convergent) sums of leading and subleading singularities respectively:
\eqn\splitsing{\eqalign{
P^{gg}_s(x,t)&=2C_A\frac{1}{x}\sum_{n=3}^{\infty}a_{n+1}\frac{1}{n!}
\lambda_s(t)^{n}\xi^{n},\cr
P^{gq}_s(x,t)&=\smallfrac{C_F}{C_A}P^{gg}_s(x,t),\qquad
P^{qg}_s(x,t)=P^{qq}_s(x,t)=0,\cr
P^{qg}_{ss}(x,t)&=\smallfrac{16}{3}\ln 2 C_A T_R n_f
\frac{1}{x}\sum_{n=1}^{\infty}
\tilde a_{n+1}\frac{1}{n!}\lambda_s(t)^{n}\xi^{n},\cr
P^{qq}_{ss}(x,t)&=\smallfrac{C_F}{C_A}P^{qg}_{ss}(x,t),\cr
P^{gg}_{ss}(x,t)&=16\ln 2 C_A^2\frac{1}{x}\sum_{n=1}^{\infty} b_{n+1}
\frac{1}{n!}
\lambda_s(t)^{n}\xi^{n},\cr
P^{gq}_{ss}(x,t)&=16\ln 2 C_A C_F
\frac{1}{x}\sum_{n=1}^{\infty} b_{n+1}^\prime\frac{1}{n!}
\lambda_s(t)^{n}\xi^{n},\cr
}}
where $\lambda_s(t)\equiv 4\ln 2\smallfrac{C_A}{\pi}\alpha_s(t)$,
$\alpha_s(t)$ is the two-loop running coupling, and
$\xi=\log\smallfrac{x_0}{x}$ for $x<x_0$, zero otherwise.

Similar expansions are taken for the singlet coefficient functions
\eqn\coeff{
C^{ij}(x,t) = \delta^{ij}\delta(1-x)
+\smallfrac{\alpha_s(t)}{2\pi}\big(C^{ij}_1(x)
+C^{ij}_s(x,t)\big)+\cdots,}
where $C^{ij}_1(x)$ are the usual two-loop coefficient functions, while
\eqn\coefsing{\eqalign{
C^{qg}_{s}(x,t)&=\smallfrac{2}{3}T_R n_f \frac{1}{x}\sum_{n=1}^{\infty}
c_n\frac{1}{(n-1)!}\lambda_s(t)^{n}\xi^{n-1},\cr
C^{qq}_{s}(x,t)&=\smallfrac{C_F}{C_A}C^{qg}_{s}(x,t),\qquad
C^{gg}_{ss}(x,t)=-C^{qg}_{ss}(x,t),\qquad
C^{gq}_{ss}(x,t)=-C^{qq}_{ss}(x,t).\cr}}
The nonsinglet splitting functions and nonsinglet coefficient function
remain the same as at two loops;
there are no singular contributions in the nonsinglet channels, except
in singular factorization
schemes~\ref\CHad{ S.~Catani \& F.~Hautmann, \PL\vyp{B315}{1993}{157},
                  \NP\vyp{B427}{1994}{475}\semi
        F.~Hautmann, to be published in the proceedings of {\it ``QCD94''},
        Montpellier, July 1994 (\NPBPS).}.

The coefficients $a_n$ in \splitsing\ may be
deduced~\ref\BFKLad{T.~Jaroszewicz,
 \PL\vyp{B116}{1982}{291}.} from the BFKL
kernel
by means of appropriate factorization
theorems~\ref\BFKLadproof{S.~Catani, F.~Fiorani and G.~Marchesini,
\PL\vyp{B234}{1990}{339}; \NP\vyp{B336}{1990}{18}\semi
  S.~Catani, F.~Fiorani, G.~Marchesini and G.~Oriani,
\NP\vyp{B361}{1991}{645}.}\nref\CCH{S.~Catani, M.~Ciafaloni and F.~Hautmann,
            \PL\vyp{B242}{1990}{97}; \NP\vyp{B366}{1991}{135};
            \PL\vyp{B307}{1993}{147}.}\nref\CHad{ S.~Catani \&
F.~Hautmann, \PL\vyp{B315}{1993}{157},
                  \NP\vyp{B427}{1994}{475}\semi
        F.~Hautmann, to be published in the proceedings of {\it ``QCD94''},
        Montpellier, July 1994 (\NPBPS).}\nref\Catrev{ S.~Catani,
DFF~207/6/94, talk given at {\it Les Rencontres de Physique de
La Vall\'ee d'Aoste}, La Thuile, 1994.}; a similar
technique for computing $\tilde a_n$ and $c_n$
has been  developed in~\refs{\CCH,\CHad,\Catrev}.
However, unlike the $a_n$, the coefficients
$\tilde a_n$ and $c_n$, being subleading, depend on
which renormalization and factorization schemes are adopted. In
ref.~\CHad\ calculations are performed in \MS, but results are also
given in a corresponding parton scheme (DIS) in which
the gluon contribution to $F_2$ is eliminated
altogether, the $c_n$ are zero and  $C^{ij}(x,t)=\delta^{ij}\delta(1-x)$.
Furthermore, the process independent singularity in the DIS quark
anomalous dimensions, which is what makes the corresponding
$\tilde a_n$ so large, can be eliminated by a suitable redefinition of
the normalization of the gluon distribution, thereby greatly
reducing the coefficients $\tilde a_n$; this  normalization (and
thus this form of the anomalous dimension) is automatically given by
off-shell factorization of collinear singularities ($Q_0$ factorization)
{}~\ref\cia{M. Ciafaloni, preprint CERN-TH/95-119 (May 1995).}.
This factorization prescription thus defines an alternative parton
scheme which we call \QDIS. It is also then
possible to construct a \QMS\ scheme, in which the anomalous
dimensions (and thus the $\tilde a_n$) are the same as those in the
\MS\ scheme, but the process independent singularity in the
coefficient functions is eliminated, reducing the $c_n$.
Finally, it is even possible to construct a parton scheme
in which both the $\tilde a_n$ and the $c_n$ are zero
(the singular DIS or SDIS scheme)~\ref\catsdis
{S. Catani, unpublished (private communication).};  all subleading
singularities in the quark channel are then factorized into the
starting distribution.

Analytic expressions for the first
fourteen coefficients $a_n$ may be found in the second of ref.~\BFKLadproof,
and for the first five $\tilde a_n$, $c_n$
in~\CHad, while the first fourteen $\tilde a_n$ and twelve $c_n$ in
\MS\  may be found in table~2 and table~3.
Analytic expressions for the $\tilde a_n$ in DIS scheme are readily
deduced from tables 2 and 3, and the corresponding expressions
in the \QDIS\ scheme may be computed similarly.
Numerical values for the first thirty-two $a_n$, and $\tilde a_n$ in
DIS are given in
\Summing; in table~1 we give also the first twenty $\tilde a_n$
and the corresponding $c_n$ in \MS, the $c_n$ in \QMS, and the
$\tilde a_n$ in \QDIS.\foot{These latter coefficients are
obtained~\cia\ in practice
by simply suppressing the singular process independent $R_N$ factors in
the expressions in ref.~\refs{\CCH-\Catrev}.}

The coefficients $b_n$ and $b_n^\prime$ which appear in the expansion of the
NLO singular splitting functions $P^{gg}_{ss}$ and $P^{gq}_{ss}$
have not yet been explicitly computed in any scheme. However, when
working at NLO in the double leading scheme no computation
is really necessary, since the requirement of
momentum conservation determines these coefficients uniquely.
Indeed, in any subtraction scheme in which there is
a detailed balance of momentum between quarks and gluons, the
splitting functions must satisfy the relations
\eqn\mom{\int_0^1\! dx\,x[P^{qg}(x,t)+P^{gg}(x,t)]=0,\qquad
\int_0^1\! dx\,x[P^{qq}(x,t)+P^{gq}(x,t)]=0,}
for all $t$. This equation cannot be satisfied  at leading order, since
$P^{qg}_s=P^{qq}_s=0$, but at NLO it
determines\eqnn\nloggcoeff\foot{Note
that if these coefficients
were computed explictly in a given renormalization and factorization
scheme the result would not
necessarily agree with eq.~\nloggcoeff\ and thus would not conserve
momentum at NLO in the double leading expansion. However, it is always
possible to then perform a change of factorization scheme such that momentum
conservation is restored at NLO, and thus eq.~\nloggcoeff\ is
satisfied.}
$$b_n = b_n^\prime = - a_{n+1}
-\smallfrac{T_R n_f}{3C_A}\tilde a_n.\eqno{\nloggcoeff}$$
This procedure can be extended to any finite order: for example
at NNLO $P^{gg}_{ss}$, $P^{gq}_{ss}$, $P^{qg}_{sss}$
and $P^{qq}_{sss}$ would be computed in \MS, and then
$P^{gg}_{sss}$ and $P^{gq}_{sss}$ would be fixed by momentum
conservation. Notice that at NLO this prescription automatically
preserves the colour-charge relation
\eqn\colch{P^{gq}_{ss}(x,t)=\smallfrac{C_F}{C_A}P^{gg}_{ss}(x,t);}
it is then easy to check
explicitly that because of the colour-charge
relations between $P^{gq}$ and $P^{gg}$, $P^{qq}$ and $P^{qg}$
the supersymmetry relation
\eqn\splitsusy{P^{qq}(x,t)+P^{gq}(x,t)=P^{qg}(x,t)+P^{gg}(x,t),}
holds when $C_F=C_A=2T_R$.

Whereas the double leading expansion scheme discussed so far is appropriate
to the HERA region, it is clearly inadequate at large $x$:
there, in order to improve the two-loop expressions of splitting
functions, it would be necessary to compute the full three loop
contributions $P^{ij}_3$ and $C^{ij}_2$, while to include only those
higher order
contributions which have small $x$ singularities is meaningless.
Below a reference value $x_0$ the ordinary loop expansion scheme
should thus be used instead, so the singular contributions $P^{ij}_s$,
$P^{ij}_{ss}$ and $C^{ij}_s$ should all vanish there.  Imposing that
at $x=x_0$ the splitting functions and coefficient functions should be
continuous, while for $x<x_0$ the logarithmic singularities are built up
asymptotically as the evolution length $x_0/x$ increases, then leads
naturally to the expressions \splitsing\ and \coefsing.\foot{
Note that in ~\refs{\DAS,\Summing} we defined the separation
of the evolution region and introduced different expansion schemes in
different regions at the level of the evolution equation, by
implicitly evolving the parton distribution functions
at large $x$ to give the boundary condition at $x=x_0$ from which they
were then evolved using an expansion scheme appropriate
to small $x$. Here the
separation is done instead at the level of the evolution kernel, thus
making the separation into large and small-$x$
regions on the basis of the evolution length (in $x$).
The present prescription has the advantage that it may be used in
Mellin space, since the evolution remains translationally invariant.
The two prescriptions however are identical by construction when
$x>x_0$, and only differ by
subleading corrections in $\xi$ when $x<x_0$.}
We will assume for simplicity that the parameter $x_0$ which sets the
transition to the small $x$ region is
independent of $t$. We then treat it as a free parameter, to be
eventually determined empirically. Naively we would expect it to lie in the
range $0.01\lsim x_0\lsim 0.1$, at least for all reasonable schemes;
here we will thus consider the extreme case $x_0=0.1$, the
opposite extreme being given in effect by the
two loop computation (for which $x_0=0$). Notice that this
parameter will in general be factorization and renormalization scheme
dependent, and should for instance
be smaller in \MS\ (which has large values of $c_n$ and
$\tilde a_n$) than in the \QMS\ scheme.

Since the expansions \splitsing\ and \coefsing\ are convergent, they may
in practice be truncated after only a finite number of terms: when
working in Mellin space arbitrarily accurate results
may thus be obtained by truncating the corresponding series expansions of the
anomalous dimensions \Summing\ (even though these
series have only a finite radius of convergence, diverging for small
$N$). Defining
$\gamma^{ij}(N,t)\equiv\int_0^1 dx x^{N} P^{ij}(N;t)$ term by term, we thus
have
\eqn\adsing{
\eqalign{
\gamma^{gg}_s(N,t)&\simeq 2C_A x_0^{-N}\sum_{n=4}^{n_{\rm max}}a_n
\lambda_s(t)^{n-1}/N^n,\cr
\gamma^{gq}_s(N,t)&=\smallfrac{C_F}{C_A}\gamma^{gg}_s(N,t),\qquad
\gamma^{qg}_s(N,t)=\gamma^{qq}_s(N,t)=0,\cr
\gamma^{qg}_{ss}(N,t)&\simeq \smallfrac{16}{3}\ln 2 C_A T_R n_f x_0^{-N}
\sum_{n=2}^{n_{\rm max}}
\tilde a_n\lambda_s(t)^{n-1}/N^{n},\cr
\gamma^{gg}_{ss}(N,t)&\simeq -16\ln 2 C_A^2 x_0^{-N}\sum_{n=2}^{n_{\rm max}}
\big(a_{n+1}+\smallfrac{T_R n_f}{3C_A}\tilde a_n\big)
\lambda_s(t)^{n-1}/N^{n},\cr
\gamma^{qq}_{ss}(N,t)&=\smallfrac{C_F}{C_A}\gamma^{qg}_{ss}(N,t),\qquad
\gamma^{gq}_{ss}(N,t)=\smallfrac{C_F}{C_A}\gamma^{gg}_{ss}(N,t).\cr
}}
The renormalization group equation may then be integrated explicitly,
while the inverse
Mellin transform is straightforward because the anomalous dimensions
\adsing\ are meromorphic in $N$: in particular there are no extra
singularities or branch cuts~\Summing.\foot{A quite complicated set of
branch cuts are present in the analytic continuation of the complete
series from large $N$~\ref\EHW{R.K.~Ellis, F.~Hautmann and B.R.~Webber,
Cavendish-HEP-94/18, Fermilab-PUB-95/006-T, {\tt hep-ph/9501307}.}.}
Care must be taken to
ensure that all the subleading  contributions are consistently
linearized in this solution (even across thresholds).  It is important
to notice that this applies to the full NLO contribution, and not only
to its large-$x$ part; the effect
of the spurious sub-subleading terms which would otherwise be produced
may in practice be quite large.


In order to actually perform the computation of $F_2^p$
we use a suitable generalization of the general procedure developed
for two loops calculations in ref.~\Mont.
We take a set of parton distribution functions
$\Delta=\{f(x,4\GeV^2): f=v,q,g,\ldots, x>10^{-2}\}$ fitted to
all available data with $Q^2\gsim 4\GeV^2$, $x\gsim 10^{-2}$ which
predates the HERA data and is thus unbiased by it: in practice we
work with the two  MRS distributions \DZP\ and \DMP\
\ref\MRSDZDM{A.D.~Martin R.G.~Roberts and W.J.~Stirling,
\PL\vyp{B306}{1993}{145}.}. We then evolve these distributions to some
new scale $Q_0^2$, and refit the resulting parton distribution functions
$\{f(x,Q_0^2): f=v,q,g,\ldots, x>10^{-2}\}$ to functions of the same
form as the original distributions, but with the small $x$ behaviour
of the singlet sea and the glue distributions now fixed to be
$x^{\lambda}$ for some $\lambda$:
\eqn\seaglue{\eqalign{
xq(x,Q_0^2) &= A_q x^{\lambda_q}(1-x)^{\eta_q}
(1+\epsilon_q x^{1/2}+\gamma_q x),\cr
xg(x,Q_0^2) &= A_g x^{\lambda_g}(1-x)^{\eta_g}
(1+\epsilon_g x^{1/2}+\gamma_g x).\cr}}
We thus  obtain a new set of
distributions $[\Delta](Q_0,\lambda)$
which when evolved up to larger values of $Q^2$ will give (almost)
precisely the same results for $x\gsim 10^{-2}$ as the original distribution
$\Delta$ (perturbative evolution being `causal' both in $t$ and $1/x$)
but different results for $x\lsim 10^{-2}$. The functions
$[\Delta](Q_0,\lambda)$ can actually be fitted to
parton distributions $\{f(x,Q_0^2)\}$ defined
 either in the \MS\ or in a parton scheme; possible differences in the results
will then depend on the approximation introduced by the choice of the
specific functional form of eq.~\seaglue.

Numerical computations can now be performed in Mellin space
using a straightforward extension to
$n_{\rm max}$ loops of the standard two loop solution
\ref\DFLM{W.~Furmanski and R.~Petronzio, \ZP\vyp{C11}{1982}{437}\semi
M.~Diemoz, F.~Ferroni, E.~Longo and G.~Martinelli,
\ZP\vyp{C39}{1988}{21}.} of the renormalizaton
group equations, with the form eq.~\adsing\ of the small-$x$
contributions to the anomalous dimensions. We have checked explicitly that
the evolved parton distributions converge as $n_{\rm max}$ is
increased: very accurate results in the HERA region may be obtained
using only the coefficients given in table~4, for all values of $x_0$.
Due to the universality of double
scaling we expect the results to be insensitive to the detailed form
of the starting distributions; the only significant constraint imposed
by the large-$x$ data is the overall normalization of the small-$x$ tail.
The data can thus be fitted by simply adjusting the
parameters $Q_0$, $\lambda_q$ and $\lambda_g$, taken within
suitable ranges.

Let us first consider the two loop case (or, equivalently,
double leading schemes with $x_0=0$) in the \MS\ scheme. We begin by
assuming for simplicity $\lambda_q=\lambda_g=\lambda$. Still, the
set of free parameters $(Q_0,\lambda)$ is somewhat redundant. To understand
how this works, we present in fig.~2a a $\chi^2$ contour plot in
the $\lambda-Q_0$ plane.
All the 46 ZEUS \ZEUS\ and 76 H1 \Hone\ bins which pass the
cut $\sigma,\rho >1$ are included in the computation of the
$\chi^2$; the cuts are imposed in order to only include small $x$ data.
The normalization of each experiment
is also fitted, in accordance with the experimental
uncertainty\foot{A useful discussion of the appropriate treatment of
normalization uncertainties may be found in~\ref\CS {J.C.~Collins and
D.E.~Soper, CTEQ-note-94-01, {\tt hep-ph/9411214}.}.}.
Clearly, a good fit can be obtained for a variety
of values of the starting scale $Q_0$;
 the valley floor then defines $\lambda$ as a function of
$Q_0$ (or alternatively, of course, the other way around).
As noted in \Mont, reasonable results at two loops may be
obtained when $\lambda (Q_0)$ vanishes at $Q_0\sim 1~\GeV$;
starting at a higher scale the best fit value increases, being
around $\lambda=-0.2$ if $Q_0\simeq2$~GeV,
consistent with the most recent MRS
global fit  \ref\MRSG{A.D.~Martin,
W.J.~Stirling and R.G.~Roberts, preprint RAL-95-021,
DTP/95/14, {\tt hep-ph/9502336.}}.
What really matters however is not
so much the precise value of $\lambda$, but that the data prefer
soft starting distributions, which necessarily give rise to
double scaling in the HERA region:
as explained in \DAS\ if the starting distribution
is too hard, say $\lambda (1~\GeV^2)\lsim -0.2$ or
$\lambda (4~\GeV^2)\lsim -0.4$, double
scaling is spoiled and a satisfactory fit to the data is no longer
obtained. Similarly the fit gets worse if one attempts to describe
the data using a power at a much higher scale.

In table~4 we show thus the best fit value of the parameter
$\lambda$ obtained in the \MS\ scheme at various values of $Q_0$.
Whereas the double scaling prediction of ref.~\DAS\ with $Q_0=1$~GeV
already
provides a good overall description of the data, the fit improves in
a statistically significant way when the full two loop calculation is
performed, especially if, consistent with the $\chi^2$ plot of
\fig\csqql{Contours of equal $\chi^2$ in the $\lambda$-$Q_0$ plane
with $\alpha_s(M_Z) = 0.120$: a) at two loops, \MS\ scheme
b) double leading DIS scheme with $x_0=0.1$. $Q_0$ is in $\GeV$.
The $\chi^2$ is computed in the same way as in table~4.}a, we choose
a somewhat larger starting scale.
The improvement in the fit is  also apparent in the corresponding two
loop curves in the scaling plot \scaling. Evolution in DIS scheme at two
loops gives almost identical results.

We now proceed to NLO calculations in the double leading
expansion scheme with $x_0=0.1$, and in the various factorization
and renormalization schemes discussed above.
The result found in the SDIS scheme (which differs from the two
loop computation DIS only for the inclusion of gluon leading singularities
$\gamma_s^{gg}$~\adsing) are effectively indistinguishable from those obtained
at two loops\foot{This is partly because the first few
$a_n$ vanish, but also because the others are very small, so that
their effect is concentrated in a very narrow wedge close to the
boundary~\Summing.}
(at least in the range of present-day data); we will thus not discuss
this case further here. If the \QMS\ scheme is used instead
the best-fit value of the parameter $\lambda$ changes appreciably;
however the impact on the quality of the fit does not appear to
be statistically significant. If we use instead the double
leading \MS\ scheme, the agreement with the data of
the present fit is significantly worse. This, however, turns out to
be a byproduct of the specific way we are parametrizing the boundary
conditions.

In order to understand this, we present in table~5 the results of fits
analogous to those discussed so far, but where now we fit the
small-$x$ exponents of the quark and gluon distributions
\seaglue, $\lambda_q$ and $\lambda_g$ as two independent parameters.
Furthermore, we fit the functional form of eq.~\seaglue\ to parton
distributions in either \MS\ or parton schemes; in order to
disentangle effects due to the fitting from the true scheme dependence
we then evolve both sets of distributions in both \MS\ and parton schemes.
It then turns out that in some schemes
(and in particular in the double leading \MS\ scheme)
a good fit is only possible if the values of $\lambda_q$
and $\lambda_g$ are substantially different.
In particular, at two loops with the fitted distribution in \MS\
$\lambda_q\simeq\lambda_g$, but
if the distribution is fitted in DIS
$\lambda_q$ is significantly lower than $\lambda_g$. In the double
leading expansion the situation is reversed:
$\lambda_q\simeq\lambda_g$ in DIS, but $\lambda_g$ is very
significantly lower than $\lambda_q$ in \MS.\foot{It is
interesting to notice that the best fit $\lambda_g$ is then
rather large; nevertheless scaling is not spoilt. This agrees
with the results of ref.~\Summing, where we show that
although at two loops only $\lambda_g\gsim -0.3$ leads to scaling,
in the double leading scheme any $\lambda_g\gsim -\lambda_s(Q_0)$
is sufficient.}
However, if the double leading calculation
is performed with  Q$_0$ factorization, then $\lambda_q\simeq\lambda_g$
in either case.
Notice that this is genuinely an effect of the fitting, in that all of
the above remains true regardless of whether the evolution is
performed in \MS\ or in a parton scheme.

We conclude that in general the results of the double leading calculations
do not differ significantly from those obtained
at two loops. This means that,  while the leading singularities
$P_s^{gg}$ have essentially no effect at all, most of the effect
of the subleading singularities  $P_{ss}^{qg}$
on $F_2$ in the HERA region may be absorbed into the
boundary condition, generally reducing $\lambda$ and increasing $Q_0$.
The HERA data on $F_2^p$ thus do not as yet actually allow us to fix
the value of $x_0$.
However, the boundary conditions themselves do change significantly
as we go from two loops to the double leading scheme; furthermore,
in the double leading scheme they display very strong scheme dependence.
This is of course the essence of factorization: since the boundary
condition at $Q_0^2$ is essentially nonperturbative and scheme
dependent, it must always be fitted in a comparison with the
data, and furthermore it may change substantially when
different factorization and renormalization schemes are adopted.
This however has also the practical consequence that we
may only reasonably take $\lambda_q=\lambda_g\equiv\lambda$ in the fitted
distribution if we fit \MS\ distributions at two loops, and
DIS distributions in the double leading scheme, unless we use
Q$_0$ factorization, in which case we may use either.
In order to avoid artifical complications related to the
fitting procedure we shall thus henceforth perform
two loop computations (fitting and, for definiteness, evolution)
in \MS\ and double leading computations in parton schemes; this will allow
us to take $\lambda_q=\lambda_g=\lambda$ throughout.

We can now explore in more detail the $Q_0$ dependence
of $\lambda$ in the
double leading DIS scheme, as summarized in the $\chi^2$ contour plot
\csqql b. The best fit $\lambda$ at a given scale is
significantly reduced in comparison with the two loop case \csqql a
because $F_2$ grows
more quickly near the boundary as $Q^2$ is increased.\foot{Very low
values of the starting scale $Q_0$ (as favoured by \ref\GRV{
M.~Gl\"uck, E.~Reya and A.~Vogt, \PL\vyp{B306}{1993}{391} and
ref. therein.}, for example) would seem to be excluded in these schemes,
because their effect when $\alpha_s(Q_0^2)$ is large is too great to
admit a satisfactory fit to the data even with valence-like input (large
positive $\lambda$).} The scale $Q_0$ at which a  given (soft) boundary
condition should be set is thus correspondingly increased~\Summing:
the results obtained then differ very little from the two loop fits, except
at large $\rho$ i.e. very close to the boundary itself,
as displayed by the pertinent curves in \scaling.
Furthermore, because of the more significant role played by perturbative
evolution, the double leading fit is more sensitive to the choice of
$Q_0$. However, the fit is less sensitive to the precise value of $\lambda$.
This somewhat surprising result was anticipated in ref.~\Summing\ where
it was shown that actually double scaling is obtained from a wider
set of initial conditions if a full double leading computation is
performed than would be the case at two loops.


It is now relatively straightforward to determine $\alpha_s$ from the
HERA data on $F_2$: we simply include an extra parameter
$\Lambda_{\rm QCD}$ (or equivalently $\alpha_s(M_Z)$) in the fit.
The results of such two parameter fits, for the various NLO calculations
described above
are displayed in table~6. The experimental uncertainty in the
resulting value of $\alpha_s(M_Z)$
turns out to be remarkably small, of the order of a few per cent.
The value of the
starting scale $Q_0$ is seen to have little effect on the
value of $\alpha_s(M_Z)$, though $\lambda(Q_0^2)$ varies rather strongly
when $Q_0$ is below 2~GeV (see \csqql).

The dependence on $Q_0$ and $\lambda$ is explored in detail in
the $\chi^2$ plots in the two
planes perpendicular to that of \csqql: the $\alpha$ - $\lambda$
plane \fig\csqal{As \csqql, but the $\alpha$ - $\lambda$ plane, with
$Q_0 = 2~\GeV$.}, and the $Q_0$-$\alpha$ plane \fig\csqqa{As \csqql,
but now the $Q_0$ - $\alpha$ plane, with a) $\lambda = -0.23$ in the
two loop calculation and b) $\lambda = -0.06$ in the double-leading
calculation (see
table~6).}. The well-defined minimum in $\lambda$ and $\alpha_s(M_Z)$
is apparent from \csqal a; large values of $\alpha_s$ are excluded
much more strongly when the subleading singularities are included
\csqal b. The flatness of the valley floor in \csqqa\ reflects that in
\csqql, but again the acceptable range of $Q_0$ is significantly
reduced when the subleading singularities are included in the calculation.
In general all double leading calculations seem to be significantly
more sensitive to the values of the parameters; even though they
cannot yet be favoured on statistical grounds they could thus
potentially lead to a firmer determination of $\alpha_s$.
For the time being, however, we determine the central value
of $\alpha_s$ from the fits with the best $\chi^2$,
namely, the two loop calculations. An average of these (see table~6)
gives a central value $\alpha_s(M_Z)=0.120$. The uncertainty related
to the possibility of choosing various expansion schemes is then
asymmetric, since double leading schemes tend to
lead to a reduction of $\alpha_s$; from table~6 we estimate it to be
$+0.002-0.006$.

We now examine all other sources of error in turn.
The sensitivity to the form of the starting
distributions is  explored by considering different choices of large $x$
distributions (for example the two `extremal' MRS distributions
\DZP\ and \DMP\ \MRSDZDM), and also more importantly by varying the way in
which the small-$x$ tail is fitted to these distributions at
$Q_0$. In particular, choosing $\lambda_s\neq\lambda_g$ has little
effect on $\alpha_s$ (see table~7). Two alternative fitting procedures
have also been tried: one in which a small-$x$ tail of the form
$x^\lambda$ is added by hand to the singlet distributions
$xq(x,Q_0^2)$, $xg(x,Q_0^2)$ before the refitting (which is then
performed over the extended range $x>10^{-4}$), the normalization of
the tail being fixed by continuity, and a similar one in which the
tail has instead the form $x^{\lambda}(1+\epsilon\sqrt{x})$, the parameters
$\epsilon_q$ and $\epsilon_g$ being chosen such that the derivatives
of the two distributions were continuous at the matching point (see
again table~7). Estimating the error from all these sources rather
conservatively at $\pm 0.003$, and combining
it in quadrature with an error of $\pm 0.003$ from
the two parameter fit, and $\pm0.002$ from the choice of $Q_0$ (see table~6),
we feel confident in quoting an overall experimental uncertainty of
$\pm 0.005$.

The most important theoretical error is, just as in the large $x$
determinations \ref\lxalf{
A.D.~Martin, W.J.~Stirling and R.G.~Roberts, \PL\vyp{B266}{1991}{173}\semi
M.~Virchaux and A.~Milsztajn,\PL\vyp{B274}{1992}{221}.}, the
uncertainty due to unknown higher order
perturbative corrections. This is manifest both in the scheme
dependence (results contrasting \MS\ and DIS schemes are presented in
table~8), and the dependence on the choice of factorization scale
$M^2\equiv k_1 Q^2$ and renormalization scale $\mu^2\equiv k_2
Q^2$ (table~8). Since at present there is insufficient data to rule out all but
very large changes in these scales empirically,\foot{With the
exception of significant reductions in the factorization scale.}
we simply varied them by a factor of two either side, as is usual in the jet
determinations~\alfrev; this gives an uncertainty of $+0.008-0.004$. It
could be argued that larger variations are already ruled out in deep inelastic
scattering by the large $x$ data \lxalf.

The position of the quark thresholds is varied from $\half m_q$ to $3
m_q$ (with $m_c=1.5\GeV$, $m_b=4.8\GeV$), to give an overall
uncertainty on $\alpha_s(M_Z)$ of $\pm 0.002$: varying the
prescription used for taking the running coupling across the thresholds
gives a further uncertainty of $\pm 0.001$.
To estimate the effect of higher twist corrections, and possible mass
effects at the quark thresholds, we perform the analysis with a
$Q^2$ cut on the data, discarding all data with either
$Q^2 > Q_{\rm cut}^2$, or $Q^2 < Q_{\rm cut}^2$. The results are
displayed in table~9, and show a remarkable level of stability as
$Q_{\rm cut}$ is varied. We can thus conclude that higher twist
effects, since they would vary rapidly with $Q^2$, must be extremely
small throughout the HERA region. Similarly any mass effects
at the quark threshold must also have only a small effect on
$\alpha_s$: most of the statistical
significance of our determination comes from the data points with the
smallest values of $x$, and thus values of $Q^2$ above the charm
threshold but below the beauty threshold.

All these various possible sources of error are summarised in table~10:
combining the theoretical errors in quadrature, we quote a value
\eqn\alphas{\alpha_s(M_Z)=0.120\pm 0.005{\rm (exp)}\pm 0.009{\rm (th)}.}
This is quite consistent with other determinations, in particular
those from fixed target deep inelastic scattering data \lxalf, and the world
average \alfrev. It is difficult to say at precisely which  value of
$Q^2$ the result \alphas\ is obtained, since the two loop running over
quite a wide range of $Q^2$ was a crucial ingredient in the match to
the data~\Test, and thus to the determination of
$\alpha_s(Q^2)$. However it should be clear from table~9 that the most
important $Q^2$ range is $4\GeV^2 \lsim Q^2 \lsim 100\GeV^2$.

To summarise, we have shown that perturbative QCD works extremely well
in the kinematic region explored so far at HERA. This remains true
even when the perturbation series is reorganised to sum all leading
and next-to-leading large logarithms, though the data are as yet
insufficient to show to what extent this is necessary. Furthermore,
since the shape of
$F_2$ is then relatively free of nonperturbative uncertainties,
varying rather rapidly with $\alpha_s$, it can be used to provide a
surprisingly precise determination of $\alpha_s$. We have performed
such a determination using the 1993 HERA data, with the result
\alphas. The experimental error, while already relatively small,
should improve substantially when the 1994 data become available, and the
expected
fivefold increase in statistics might also provide an empirical
determination of the as yet unknown parameter $x_0$, and enable a
reduction in the dominant theoretical error coming from the
renormalization scale dependence. Although the present determination
has much in common with jet determinations, it has the advantage that
it is sufficiently inclusive that there are no uncertainties from
hadronization, and in fact here higher twist effects seem to be very small
indeed. Also, the summation of large logarithms is relatively easy to
control, essentially due again to the inclusive nature of the
measurement, but also to some accidental cancellations. This means
that a further reduction in the theoretical
error will presumably be possible with a calculation of the three
loop anomalous dimensions for deep inelastic scattering, and of the
subleading singularities at small $x$.

\bigskip
{\bf Acknowledgements}: We would like to thank G.~Altarelli,
R.K.~Ellis, R.G.Roberts,
S.~Catani, M.~Ciafaloni, A.~Cooper-Sarkar,
F.~Hautmann, R.~Nisius and
S.~Bethke for various discussions during the course of this work.

\vfill\eject
\listrefs
\vfill\eject
\listfigs
\vfill
\eject
\nopagenumbers

\midinsert\hfil
       \vbox{
      \baselineskip\footskip
     {\tabskip=0pt \offinterlineskip
      \def\tablerule{\noalign{\hrule}}
      \halign to 400pt{\strut#&\vrule#\tabskip=1em plus2em
                   &#\hfil&\vrule#
                   &#\hfil&\vrule#
                   &#\hfil&#
                   &#\hfil&\vrule#
                   &#\hfil&#
                   &#\hfil&\vrule#\tabskip=0pt\cr\tablerule
             &&\omit\hidewidth $n$\hidewidth
             &&\omit\hidewidth $\tilde a_n$[\MS] \hidewidth
             &&\omit\hidewidth $c_n$[\QMS]  \hidewidth
             &&\omit\hidewidth $c_n$[\MS] \hidewidth
             &&\omit\hidewidth $\tilde a_n$[\QDIS] \hidewidth
             &&\omit\hidewidth $\tilde a_n$[DIS]\hidewidth
              &\cr\tablerule
&&   0  && 1          && 1          && 1          && 1          && 1 &
\cr
&&   1  && 0.60112293 && 0.53664997 && 0.53664997 && 0.78145981 && 0.78145981 &
\cr
&&   2  && 0.20235532 && 0.42906025 && 1.26303490 && 0.29913310 && 0.29913310 &
\cr
&&   3  && 0.16029498 && 0.19506502 && 0.77061130 && 0.23767037 && 0.38806675 &
\cr
&&   4  && 0.09325224 && 0.31745293 && 0.66181242 && 0.14877117 && 0.25256339 &
\cr
&&   5  && 0.04811765 && 0.18492007 && 0.78619278 && 0.11628239 && 0.17838311 &
\cr
&&   6  && 0.05656923 && 0.14537345 && 0.53159913 && 0.10092750 && 0.22632345 &
\cr
&&   7  && 0.03508260 && 0.16442189 && 0.51841609 && 0.07337509 && 0.15473331 &
\cr
&&   8  && 0.02509704 && 0.10681121 && 0.55225420 && 0.06616794 && 0.13891462 &
\cr
&&   9  && 0.02919748 && 0.10018804 && 0.41447382 && 0.05783445 && 0.15744938 &
\cr
&&  10  && 0.01903403 && 0.10015474 && 0.42612176 && 0.04649840 && 0.11602071 &
\cr
&&  11  && 0.01660674 && 0.07307916 && 0.42532669 && 0.04371716 && 0.11532868 &
\cr
&&  12  && 0.01780682 && 0.07253155 && 0.34787922 && 0.03845962 && 0.11997040 &
\cr
&&  13  && 0.01250523 && 0.06820181 && 0.36161794 && 0.03308122 && 0.09568020 &
\cr
&&  14  && 0.01206009 && 0.05472576 && 0.34791451 && 0.03142577 && 0.09846280 &
\cr
&&  15  && 0.01202880 && 0.05482194 && 0.30399755 && 0.02796813 && 0.09707906 &
\cr
&&  16  && 0.00917087 && 0.05018804 && 0.31366444 && 0.02513443 && 0.08291446 &
\cr
&&  17  && 0.00919550 && 0.04317398 && 0.29683046 && 0.02389106 && 0.08559381 &
\cr
&&  18  && 0.00874215 && 0.04291270 && 0.27186534 && 0.02157069 && 0.08203738 &
\cr
&&  19  && 0.00717749 && 0.03903673 && 0.27677753 && 0.01992506 && 0.07384063 &
\cr
\tablerule}}}
\hfil\bigskip
\centerline{\vbox{\hsize= 380pt \raggedright\noindent\footnotefont
Table 1: Numerical values of the first twenty coefficients $\tilde a_n$ and
$c_n$ in the $Q_0$ schemes and \MS\ schemes. The $\tilde a_n$ in \QMS\
are the same as those in \MS\ by construction, while the $\tilde a_n$
in DIS were given in ref.\Summing.
}}
\bigskip
\endinsert
\vfill
\eject
\vfill
\eject
\midinsert\hfil
\vbox{
      \baselineskip\footskip
     {\tabskip=0pt \offinterlineskip
      \def\tablerule{\noalign{\hrule}}
      \halign to 365pt{\strut#&\vrule#\tabskip=1em plus2em
                   &\hfil#\hfil&\vrule#
                   &#\hfil&\vrule#
                   &\hfil#&\vrule#\tabskip=0pt\cr\tablerule
             &&\omit\hidewidth $n$\hidewidth
             &&\omit\hidewidth $\tilde a_n\times(4\ln 2)^n$ \hidewidth
              &\cr\tablerule
&&   0 && $  1 $ & \cr
&&   1 && $  {5\over 3}  $ & \cr
&&   2 && $  {{14}\over 9}  $ & \cr
&&   3 && $  {{82}\over {81}} + 2\,\zeta_3  $ & \cr
&&   4 && $  {{122}\over {243}} + {{25\over 6} \,\zeta_3}$ & \cr
&&   5 && $  {{146}\over {729}} + {{14}\over 3}\,\zeta_3 +
             2\,\zeta_5  $  & \cr
&&   6 && $  {{2188}\over {32805}} + {{287\over {81}}\,\zeta_3} +
   12\,{{\zeta_3}^2} + {{35}\over 9} \,\zeta_5$
& \cr
&&   7 &&$   {{13124}\over {688905}} + {{488\over {243}}\,\zeta_3} +
   {{515\over {21}}\,{{\zeta_3}^2}} +
   {{112}\over {27}}\,\zeta_5 + 2\,\zeta_7  $
& \cr
&&   8 &&$   {{1406}\over {295245}} + {{73\over {81}}\,\zeta_3} +
   {{55}\over 2}\,{{\zeta_3}^2} + {{82}\over {27}}\,\zeta_5 +
   32\,\zeta_3\,\zeta_5 + {{15\over 4}\,\zeta_7}$
& \cr
&&  9 &&$ {{11810}\over {11160261}} + {{2188\over {6561}}\,\zeta_3} +
   {{2665\over {126}}\,{{\zeta_3}^2}} + 96\,{{\zeta_3}^3} +
   {{1220\over {729}}\,\zeta_5} +
   {{1675}\over {27}}\,\zeta_3\,\zeta_5 +
   {{35}\over 9}\,\zeta_7 + 2\,\zeta_9$& \cr
&&  10 && $ {{177148}\over {837019575}} + {{72182}\over {688905}}\,\zeta_3 +
   {{20801\over {1701}}\,{{\zeta_3}^2}} +
   {{1342\over 7}\,{{\zeta_3}^3}} +{{1606}\over {2187}}\,\zeta_5 +
   {{5390}\over {81}}\,\zeta_3\,\zeta_5 +
  $
& \cr
&&  && \qquad   $ 20\,{{\zeta_5}^2} + {{451}\over {162}}\,\zeta_7 +
   40\,\zeta_3\,\zeta_7 + {{11}\over 3}\,\zeta_9$ & \cr
&&  11 &&${{1062884}\over {27621645975}} + {{2812}\over {98415}}\,\zeta_3 +
   {{9563}\over {1701}}\,{{\zeta_3}^2} +
   {{9539}\over {45}}\,{{\zeta_3}^3} +
   {{8752}\over {32805}}\,\zeta_5 +{{11972}\over {243}}\,\zeta_3\,\zeta_5 +
$ & \cr
&&    && \qquad$      440\,{{\zeta_3}^2}\,\zeta_5 +
   {{3710}\over {99}}\,{{\zeta_5}^2} +
   {{122}\over {81}}\,\zeta_7 +
   {{1655}\over {22}}\,\zeta_3\,\zeta_7 +
   {{56}\over {15}}\,\zeta_9+ 2\,\zeta_{11}$& \cr
&&  12 && ${{1594324}\over {248594813775}} +
   {{76765}\over {11160261}}\,\zeta_3 +
   {{163553}\over {76545}}\,{{\zeta_3}^2} +
   {{1835119}\over {11340}}\,{{\zeta_3}^3} + 880\,{{\zeta_3}^4} + $
 & \cr
&&  && $\qquad
   {{170612}\over {2066715}}\,\zeta_5 +
{{60268}\over {2187}}\,\zeta_3\,\zeta_5 +
   {{159770}\over {189}}\,{{\zeta_3}^2}\,\zeta_5 +
   {{34762\over {891}}\,{{\zeta_5}^2}} +
   {{949}\over {1458}}\,\zeta_7 + $
 & \cr
&&   &&$\qquad\qquad
   {{1729}\over {22}}\,\zeta_3\,\zeta_7 +
   48\,\zeta_5\,\zeta_7 + {{1066}\over {405}}\,\zeta_9 +
 48\,\zeta_3\,\zeta_9 + {{65}\over {18}}\,\zeta_{11}$ & \cr
&&  13 &&$ {{1913188}\over {1939039547445}} +
   {{177148}\over {119574225}}\,\zeta_3 +
   {{95149}\over {137781}}\,{{\zeta_3}^2} +
   {{113399}\over {1215}}\,{{\zeta_3}^3} +
   {{471136}\over {273}}\,{{\zeta_3}^4} + $ & \cr
&&  && $ \qquad{{19684}\over {885735}}\,\zeta_5 +
   {{80738}\over {6561}}\,\zeta_3\,\zeta_5 +
   {{73115}\over {81}}\,{{\zeta_3}^2}\,\zeta_5 +
   {{75194}\over {2673}}\,{{\zeta_5}^2} +
   624\,\zeta_3\,{{\zeta_5}^2} +$
& \cr
&&   &&$\qquad\qquad     {{7658}\over {32805}}\,\zeta_7 +
   {{101311}\over {1782}}\,\zeta_3\,\zeta_7 +  624\,{{\zeta_3}^2}\,\zeta_7 +
   {{20615}\over {234}}\,\zeta_5\,\zeta_7 +
   {{1708}\over {1215}}\,\zeta_9 + $  & \cr
&&   &&\qquad\qquad\qquad$    {{3451}\over {39}}\,\zeta_3\,\zeta_9 +
   {{98}\over {27}}\,\zeta_{11} + 2\,\zeta_{13}$ & \cr
\tablerule}}}
\hfil\bigskip
\centerline{\vbox{\hsize= 380pt \raggedright\noindent\footnotefont
Table 2: Exact expressions for the first fourteen coefficients $\tilde a_n$
in \MS\ scheme.
}}
\bigskip
\endinsert
\vfill
\eject
\midinsert\hfil
\vbox{
      \baselineskip\footskip
     {\tabskip=0pt \offinterlineskip
      \def\tablerule{\noalign{\hrule}}
      \halign to 380pt{\strut#&\vrule#\tabskip=1em plus2em
                   &\hfil#\hfil&\vrule#
                   &#\hfil&\vrule#
                   &\hfil#&\vrule#\tabskip=0pt\cr\tablerule
             &&\omit\hidewidth $n$\hidewidth
             &&\omit\hidewidth $c_n\times(4\ln 2)^n$ \hidewidth
              &\cr\tablerule
&&   0 && $ 1$   & \cr
&&   1 && $ {{43}\over 9} - {{{{\pi }^2}}\over 3}$   & \cr
&&   2 && $ {{1234}\over {81}} - {{13\,{{\pi }^2}}\over {18}} +
   {{4}\over 3}\,\zeta_3$& \cr
&&   3 && $  {{7412}\over {243}} - {{71\,{{\pi }^2}}\over {54}} -
   {{2\,{{\pi }^4}}\over {15}} + {{89}\over 9}\,\zeta_3 $ & \cr
&&   4 && $ {{50012}\over {729}} - {{233\,{{\pi }^2}}\over {81}} -
     {{13\,{{\pi }^4}}\over {45}} + {{910}\over {27}}\,\zeta_3 -
     {{14\,{{\pi }^2}}\over 9}\,\zeta_3 + {{24}\over 5}\,\zeta_5  $& \cr
&&   5 && $  {{4129144}\over {32805}} - {{1276\,{{\pi }^2}}\over {243}} -
   {{71\,{{\pi }^4}}\over {135}} - {{221\,{{\pi }^6}}\over {11340}} +
   {{9074}\over {81}}\,\zeta_3 -
   {{130\,{{\pi }^2}}\over {27}}\,\zeta_3 +
   {{178}\over 9}\,{{\zeta_3}^2} + {{808}\over {45}}\,\zeta_5 $ & \cr
&&   6 &&$ {{190112792}\over {688905}} - {{8384\,{{\pi }^2}}\over {729}} -
   {{466\,{{\pi }^4}}\over {405}} - {{2873\,{{\pi }^6}}\over {68040}} +
   {{65864}\over {243}}\,\zeta_3 -
   {{923\,{{\pi }^2}}\over {81}}\,\zeta_3 -
   {{358\,{{\pi }^4}}\over {315}}\,\zeta_3 +
   $ & \cr
&&   && $\qquad {{17939}\over {189}}\,{{\zeta_3}^2}+
  {{6536}\over {135}}\,\zeta_5 -
   {{32\,{{\pi }^2}}\over {15}}\,\zeta_5 + {{60}\over 7}\,\zeta_7$   & \cr
&&   7 &&$ {{148801028}\over {295245}} - {{45928\,{{\pi }^2}}\over {2187}} -
   {{2552\,{{\pi }^4}}\over {1215}} - {{15691\,{{\pi }^6}}\over {204120}} -
   {{487\,{{\pi }^8}}\over {272160}} +
         {{535846}\over {729}}\,\zeta_3 - $ & \cr
&&    &&   $\qquad
      {{7456\,{{\pi }^2}}\over {243}}\,\zeta_3 -
      {{2873\,{{\pi }^4}}\over {945}}\,\zeta_3 +
      {{24988}\over {81}}\,{{\zeta_3}^2}
       -{{365\,{{\pi }^2}}\over {27}}\,{{\zeta_3}^2} +
   {{57896}\over {405}}\,\zeta_5 - {{91\,{{\pi }^2}}\over {15}}\,\zeta_5 +
               $& \cr
&&  &&$\qquad\qquad
   {{491}\over 5}\,\zeta_3\,\zeta_5 +
   {{1109}\over {42}}\,\zeta_7$& \cr
&&  8 && $ {{12320913596}\over {11160261}} -
        {{301808\,{{\pi }^2}}\over {6561}} -
   {{16768\,{{\pi }^4}}\over {3645}} - {{51493\,{{\pi }^6}}\over {306180}} -
   {{6331\,{{\pi }^8}}\over {1632960}} +
   {{646232}\over {405}}\,\zeta_3 -
   $  & \cr
&&   &&$ \qquad
   {{48488\,{{\pi }^2}}\over {729}}\,\zeta_3 -
   {{18673\,{{\pi }^4}}\over {2835}}\,\zeta_3 -
   {{4213\,{{\pi }^6}}\over {17010}}\,\zeta_3 +
   {{1796639\over {1701}}\,{{\zeta_3}^2}} -
   {{7241\,{{\pi }^2}}\over {162}}\,{{\zeta_3}^2} +$& \cr
&&   &&  $\qquad\qquad
   {{18620}\over {81}}\,{{\zeta_3}^3} +
   {{1189456}\over {3645}}\,\zeta_5 -
   {{1846\,{{\pi }^2}}\over {135}}\,\zeta_5 -
   {{926\,{{\pi }^4}}\over {675}}\,\zeta_5 +
   {{98863}\over {270}}\,\zeta_3\,\zeta_5 +$ & \cr
&&  && $ \qquad\qquad\qquad
   {{1340}\over {21}}\,\zeta_7 -{{58\,{{\pi }^2}}\over {21}}\,\zeta_7 +
                   {{112}\over 9}\,\zeta_9 $   & \cr
&& 9  &&$ {{1687431562504}\over {837019575}} -
   {{1653376\,{{\pi }^2}}\over {19683}} -
   {{91856\,{{\pi }^4}}\over {10935}} - {{70499\,{{\pi }^6}}\over {229635}} -
   {{34577\,{{\pi }^8}}\over {4898880}} -
   {{2861\,{{\pi }^{10}}}\over {29937600}} +$
   & \cr
&&  &&$\qquad
   {{398373436}\over {98415}}\,\zeta_3 -
   {{368896\,{{\pi }^2}}\over {2187}}\,\zeta_3 -
   {{28426\,{{\pi }^4}}\over {1701}}\,\zeta_3 -
   {{15847\,{{\pi }^6}}\over {25515}} \,\zeta_3+
   {{13980418}\over {5103}} \,{{\zeta_3}^2}-$  & \cr
&&  &&$ \qquad\qquad
   {{55735\,{{\pi }^2}}\over {486}}\,{{\zeta_3}^2} -
   {{53756\,{{\pi }^4}}\over {4725}}\,{{\zeta_3}^2}+
   {{243592}\over {243}}\,{{\zeta_3}^3} +
   {{9349244}\over {10935}}\,\zeta_5 -
   {{14446\,{{\pi }^2}}\over {405}}\,\zeta_5 - $ & \cr
&&   &&  $\qquad\qquad\quad
   {{7189\,{{\pi }^4}}\over {2025}}\,\zeta_5 +
   {{833453}\over {810}}\,\zeta_3\,\zeta_5 -
   {{4001\,{{\pi }^2}}\over {90}}\,\zeta_3\,\zeta_5 +
   {{2234\over {25}}\,{{\zeta_5}^2}} +
   {{33265}\over {189}}\,\zeta_7 -$   & \cr
&&   && $ \qquad\qquad\qquad
   {{52\,{{\pi }^2}}\over 7}\,\zeta_7 +
   {{6612}\over {35}}\,\zeta_3\,\zeta_7 +
   {{944}\over {27}}\,\zeta_9$& \cr
&&  10 && $ {{121977196479032}\over {27621645975}} -
   {{10865024\,{{\pi }^2}}\over {59049}} -
   {{603616\,{{\pi }^4}}\over {32805}} -
   {{463216\,{{\pi }^6}}\over {688905}} -
   {{113471\,{{\pi }^8}}\over {7348320}} -$ & \cr
&&   &&$\quad
   {{37193\,{{\pi }^{10}}}\over {179625600}} +
   {{2480052752}\over {295245}}\,\zeta_3 -
   {{2296400\,{{\pi }^2}}\over {6561}}\,\zeta_3 -
   {{885544\,{{\pi }^4}}\over {25515}}\,\zeta_3 -
   {{393269\,{{\pi }^6}}\over {306180}}\,\zeta_3 -$& \cr
&&   &&$\qquad
   {{100127\,{{\pi }^8}}\over {3207600}}\,\zeta_3 +
   {{123059534}\over {15309}}\,{{\zeta_3}^2} -
   {{244417\,{{\pi }^2}}\over {729}}\,{{\zeta_3}^2} -
   {{468364\,{{\pi }^4}}\over {14175}}\,{{\zeta_3}^2} +$ & \cr
&&  &&$ \qquad\quad
   {{11643262}\over {3645}}\,{{\zeta_3}^3} -
   {{33466\,{{\pi }^2}}\over {243}}\,{{\zeta_3}^3} +
   {{59513936}\over {32805}}\,\zeta_5 -
   {{10208\,{{\pi }^2}}\over {135}}\,\zeta_5 -
   {{45653\,{{\pi }^4}}\over {6075}}\,\zeta_5 -$ & \cr
&&   && $\qquad\qquad
   {{43843\,{{\pi }^6}}\over {155925}}\,\zeta_5 +
   {{780635}\over {243}}\,\zeta_3\,\zeta_5 -
   {{1625\,{{\pi }^2}}\over {12}}\,\zeta_3\,\zeta_5 +
   {{89374}\over {55}}\,{{\zeta_3}^2}\,\zeta_5 +
   {{243391}\over {825}}\,{{\zeta_5}^2} +  $& \cr
&&  && $\qquad\qquad\quad
   {{656708}\over {1701}}\,\zeta_7 -
   {{3053\,{{\pi }^2}}\over {189}}\,\zeta_7 -
   {{626\,{{\pi }^4}}\over {385}}\,\zeta_7 +
   {{2125594 }\over {3465}}\,\zeta_3\,\zeta_7 +
   {{32156}\over {405}}\,\zeta_9 -$ & \cr
&&  && $ \qquad\qquad\qquad
   {{92\,{{\pi }^2}}\over {27}}\,\zeta_9 +
   {{180}\over {11}}\,\zeta_{11}$& \cr
&&  11 &&$  {{2004668634968152}\over {248594813775}} -
   {{59521408\,{{\pi }^2}}\over {177147}} -
   {{3306752\,{{\pi }^4}}\over {98415}} -
   {{2537522\,{{\pi }^6}}\over {2066715}} -
   {{155353\,{{\pi }^8}}\over {5511240}} -$& \cr
&&   &&$ \quad
   {{203131\,{{\pi }^{10}}}\over {538876800}} +
   {{242491\,{{\pi }^{12}}}\over {91945854000}} +
   {{25554524638}\over {1240029}}\,\zeta_3 -
   {{16901248\,{{\pi }^2}}\over {19683}}\,\zeta_3 -
   {{6522752\,{{\pi }^4}}\over {76545}}\,\zeta_3 -$& \cr
&&   &&$  \qquad
   {{103219\,{{\pi }^6}}\over {32805}}\,\zeta_3 -
   {{2539043\,{{\pi }^8}}\over {33679800}}\,\zeta_3 +
   {{4336551098}\over {229635}}\,{{\zeta_3}^2} -
   {{1721324\,{{\pi }^2}}\over {2187}}\,{{\zeta_3}^2} -$& \cr
&&   &&$ \qquad\quad
   {{471014\,{{\pi }^4}}\over {6075}}\,{{\zeta_3}^2} -
   {{499339\,{{\pi }^6}}\over {170100}}\,{{\zeta_3}^2} +
   {{1697032087}\over {153090}}\,{{\zeta_3}^3} -
   {{340262\,{{\pi }^2}}\over {729}}\,{{\zeta_3}^3} + $& \cr
&&   &&$  \qquad\qquad
   {{656023}\over {243}}\,{{\zeta_3}^4} +
   {{1336379576}\over {295245}}\,\zeta_5 -
   {{687488\,{{\pi }^2}}\over {3645}}\,\zeta_5 -
   {{341578\,{{\pi }^4}}\over {18225}}\,\zeta_5 -$& \cr
&&   &&$  \qquad\qquad\quad
   {{185419\,{{\pi }^6}}\over {267300}}\,\zeta_5 +
   {{85825556}\over {10935}}\,\zeta_3\,\zeta_5 -
   {{531719\,{{\pi }^2}}\over {1620}}\,\zeta_3\,\zeta_5 -
   {{464089\,{{\pi }^4}}\over {14175}}\,\zeta_3\,\zeta_5 +$& \cr
&&   &&$  \qquad\qquad\qquad
   {{8960074}\over {1485}}\,{{\zeta_3}^2}\,\zeta_5 +
   {{17010553}\over {22275}}\,{{\zeta_5}^2} -
   {{819\,{{\pi }^2}}\over {25}}\,{{\zeta_5}^2} +
   {{5027845}\over {5103}}\,\zeta_7 - $& \cr
&&   &&$\qquad\qquad\qquad\quad
   {{23300\,{{\pi }^2}}\over {567}}\,\zeta_7 -
   {{14209\,{{\pi }^4}}\over {3465}}\,\zeta_7 +
   {{2334799}\over {1485}}\,\zeta_3\,\zeta_7 -
   {{3034\,{{\pi }^2}}\over {45}}\,\zeta_3\,\zeta_7 +$& \cr
&&    &&$ \qquad\qquad\qquad\qquad
   {{6130}\over {21}}\,\zeta_5\,\zeta_7 +
   {{254824}\over {1215}}\,\zeta_9 -
   {{715\,{{\pi }^2}}\over {81}}\,\zeta_9 +
   {{8414}\over {27}}\,\zeta_3\,\zeta_9 +
   {{4313}\over {99}}\,\zeta_{11}        $   & \cr
&&   &&   & \cr
\tablerule}}}
\hfil\bigskip
\centerline{\vbox{\hsize= 380pt \raggedright\noindent\footnotefont
Table 3: Exact expressions for the first twelve coefficients $c_n$
in \MS\ scheme.
}}
\bigskip
\endinsert
\midinsert\hfil
\vbox{\tabskip=0pt \offinterlineskip
      \def\tablerule{\noalign{\hrule}}
      \halign to 280pt{\strut#&\vrule#\tabskip=1em plus2em
                   &\hfil#\hfil&\vrule#
                   &#\hfil&\vrule#
                   &\hfil#&\vrule#
                   &\hfil#&\vrule#\tabskip=0pt\cr\tablerule
      &&\omit&&\omit\hidewidth norm.\hidewidth
             &&\omit\hidewidth $\lambda$\hidewidth
             &&\omit\hidewidth $\chi^2$\hidewidth&\cr\tablerule
   &&   a)  && $96\%\qquad 104\%$ && $0$ && $71.0$ &\cr\tablerule
   &&       && $96\%\qquad 103\%$ && $1.05\pm 0.85$  && $64.8$ &\cr
   &&   b)  && $93\%\qquad 100\%$ && $-0.25\pm 0.13$ && $59.7$ &\cr
   &&       && $95\%\qquad 101\%$ && $-0.27\pm 0.12$ && $58.6$ &\cr\tablerule
   &&   c)  && $94\%\qquad 100\%$ && $-0.26\pm 0.12$ && $59.5$ &\cr
   &&       && $94\%\qquad 100\%$ && $-0.26\pm 0.30$ && $59.6$ &\cr\tablerule
   &&   d)  && $96\%\qquad 102\%$ && $0.08\pm 0.12$  && $63.2$ &\cr
   &&       && $97\%\qquad 103\%$ && $-0.13\pm 0.17$ && $59.9$ &\cr\tablerule
   &&   e)  && $98\%\qquad 100\%$ && $0.32\pm 0.08$  && $132$  &\cr
   &&       &&$104\%\qquad 107\%$ && $-0.10\pm 0.21$ && $137$  &\cr\tablerule
}}\hfil\bigskip
\centerline{\vbox{\hsize= 380pt \raggedright\noindent\footnotefont
Table~4: The fitted normalizations to the two experiments (H1 and
ZEUS, with uncertainties of $\pm 5\%$ and $\pm 4\%$
respectively \CS), the fitted values of $\lambda$
and the associated $\chi^2$s (for $121$ degrees of
freedom). All data points with $\sigma,\rho > 1$ are included in the
fits: this cut effectively excludes all points not inside the
double scaling region. Statistical and systematic
errors for each data point have been simply combined in quadrature.
The different cases considered are a) the simple one loop double
scaling calculation of ref.~\DAS, with $\lambda$
 fixed ($Q_0=1$~GeV); b) two loops (\MS\ scheme),
$Q_0=1$~GeV, $Q_0=2$~GeV and $Q_0=3$~GeV; c) two loops, DIS scheme,
and double leading, SDIS scheme, both $Q_0=2$~GeV; d) double leading
 \QMS\ scheme, $Q_0=2$~GeV and $Q_0=3$~GeV; e) as d), but \MS\
scheme. All entries b)-e) are full NLO computations with
$\alpha_s(M_Z)=0.120$;  the double leading
calculations have $x_0 = 0.1$.}}
\bigskip
\endinsert
\vfill
\eject

\midinsert\hfil
\vbox{\tabskip=0pt \offinterlineskip
      \def\tablerule{\noalign{\hrule}}
      \halign to 350pt{\strut#&\vrule#\tabskip=1em plus2em
                   &\hfil#\hfil&\vrule#
                   &#\hfil&\vrule#
                   &\hfil#&\vrule#
                   &\hfil#&\vrule#
                   &\hfil#&\vrule#\tabskip=0pt\cr\tablerule
      &&\omit&&\omit\hidewidth norm.\hidewidth
             &&\omit\hidewidth $\lambda_q$\hidewidth
             &&\omit\hidewidth $\lambda_g$\hidewidth
             &&\omit\hidewidth $\chi^2$\hidewidth&\cr\tablerule
   &&       && $93\%\qquad 100\%$ && $-0.24\pm 0.13$ && $-0.33\pm 0.48$
                                                          && $59.7$ &\cr
   &&   a)   && $96\%\qquad 101\%$ && $-0.12\pm 0.02$ && $-0.16\pm 0.10$
                                                          && $59.8$ &\cr
   &&       && $95\%\qquad 101\%$ && $0.01\pm 0.04$ && $-0.76\pm 0.06$
                                         && $68.8$ &\cr\tablerule
   &&       && $94\%\qquad 101\%$ && $-0.24\pm 0.09$ && $-0.52\pm 0.23$
                                                          && $59.2$ &\cr
   &&   b)   && $97\%\qquad 104\%$ && $-0.12\pm 0.02$ && $-0.01\pm 0.20$
                                                          && $62.5$ &\cr
   &&       && $97\%\qquad 103\%$ && $0.10\pm 0.06$ && $0.01\pm 0.37$
                                         && $73.1$ &\cr\tablerule
   &&       && $96\%\qquad 103\%$ && $-0.23\pm 0.05$ && $0.10\pm 0.07$
                                                          && $57.3$ &\cr
   &&   c)  && $96\%\qquad 102\%$ && $-0.25\pm 0.02$ && $0.03\pm 0.16$
                                                          && $64.5$ &\cr
   &&       && $96\%\qquad 102\%$ && $-0.26\pm 0.02$ && $0.12\pm 0.17$
                                         && $72.6$ &\cr\tablerule
   &&       && $95\%\qquad 102\%$ && $-0.23\pm 0.05$ && $0.14\pm 0.10$
                                                          && $57.8$ &\cr
   &&   d)  && $95\%\qquad 101\%$ && $-0.25\pm 0.02$ && $0.02\pm 0.05$
                                                          && $62.1$ &\cr
   &&       && $96\%\qquad 103\%$ && $-0.26\pm 0.02$ && $-0.07\pm 0.10$
                                         && $68.1$ &\cr\tablerule }}
\hfil\bigskip
\centerline{\vbox{\hsize= 380pt \raggedright\noindent\footnotefont
Table~5: As table~4, but allowing $\lambda_q\not=\lambda_g$, and with
various fitting and evolution procedures.
The different cases considered are:
a) \MS\ distributions evolved in \MS; b) \MS\ distributions
evolved in DIS; c) DIS ditributions evolved in DIS; d) DIS
distributions evolved in \MS. For each entry the three cases
correspond respectively to two loops,  double leading Q$_0$ schemes,
double leading, standard schemes.
All calculations are done with
$\alpha_s(M_Z)=0.120$;  the two loop calculations have $Q_0 = 2.0$, and
the double leading calculations $Q_0 = 3~\GeV$ and $x_0 = 0.1$.}}
\bigskip
\endinsert

\midinsert\hfil
\vbox{\tabskip=0pt \offinterlineskip
      \def\tablerule{\noalign{\hrule}}
      \halign to 350pt{\strut#&\vrule#\tabskip=1em plus2em
                   &\hfil#\hfil&\vrule#
                   &#\hfil&\vrule#
                   &\hfil#&\vrule#
                   &#\hfil&\vrule#
                   &\hfil#&\vrule#\tabskip=0pt\cr\tablerule
      &&\omit&&\omit\hidewidth norm.\hidewidth
             &&\omit\hidewidth $\lambda$\hidewidth
             &&\omit\hidewidth $\alpha_s(M_Z)$\hidewidth
             &&\omit\hidewidth $\chi^2$\hidewidth&\cr\tablerule
   &&       && $93\%\qquad 100\%$
              && $0.42\pm 0.56$ && $0.1183\pm 0.0092$ && $61.4$ &\cr
   &&   a)  && $95\%\qquad 102\%$
              && $-0.23\pm 0.03$ && $0.1215\pm 0.0017$ && $58.4$ &\cr
   &&       && $96\%\qquad 103\%$
              && $-0.26\pm 0.02$ && $0.1212\pm 0.0017$ && $57.8$ &\cr\tablerule
   &&   b)  && $94\%\qquad 99\%$
              && $0.04\pm 0.10$ && $0.1186\pm 0.0043$ && $64.2$ &\cr
   &&       && $97\%\qquad 103\%$
              && $-0.12\pm 0.04$ && $0.1200\pm 0.0026$ && $59.9$ &\cr\tablerule
   &&   c)  && $93\%\qquad 100\%$
              && $-0.08\pm 0.05$ && $0.1144\pm 0.0034$ && $62.5$ &\cr
   &&       && $91\%\qquad  98\%$
              && $-0.26\pm 0.02$ && $0.1106\pm 0.0019$ && $65.7$ &\cr\tablerule
   &&   d)  && $93\%\qquad 99\%$
              && $-0.06\pm 0.06$ && $0.1140\pm 0.0040$ && $64.9$ &\cr
   &&       && $91\%\qquad 98\%$
              && $-0.26\pm 0.02$ && $0.1098\pm 0.0040$ && $67.8$ &\cr\tablerule
}}\hfil\bigskip
\centerline{\vbox{\hsize= 380pt \raggedright\noindent\footnotefont
Table~6: As table~4, but with $\alpha_s(M_Z)$ now included in the fit:
a) two loops, with $Q_0 = 1\GeV$, $2\GeV$, and $3\GeV$
respectively; b) double leading, \QMS\ scheme with $Q_0 = 2\GeV$ and
$3\GeV$; c) as b), but \QDIS\ scheme respectively;
 d) as b), but DIS scheme.
}}
\bigskip
\endinsert

\midinsert\hfil
\vbox{\tabskip=0pt \offinterlineskip
      \def\tablerule{\noalign{\hrule}}
      \halign to 350pt{\strut#&\vrule#\tabskip=1em plus2em
                   &\hfil#\hfil&\vrule#
                   &#\hfil&\vrule#
                   &\hfil#&\vrule#
                   &#\hfil&\vrule#
                   &\hfil#&\vrule#\tabskip=0pt\cr\tablerule
      &&\omit&&\omit\hidewidth norm.\hidewidth
             &&\omit\hidewidth $\lambda$\hidewidth
             &&\omit\hidewidth $\alpha_s(M_Z)$\hidewidth
             &&\omit\hidewidth $\chi^2$\hidewidth&\cr\tablerule
   &&   a)  && $95\%\qquad 102\%$
              && $-0.23\pm 0.03$ && $0.1215\pm 0.0017$ && $58.4$ &\cr\tablerule
   &&   b)  && $96\%\qquad 102\%$
              && $-0.23\pm 0.01$ && $0.1224\pm 0.0018$ && $57.8$ &\cr\tablerule
   &&   c)  && $96\%\qquad 103\%$
              && $-0.26\pm 0.03$ && $0.1224\pm 0.0016$ && $58.1$ &\cr
   &&       && $95\%\qquad 101\%$
              && $-0.22\pm 0.03$ && $0.1211\pm 0.0018$ && $59.1$ &\cr\tablerule
   &&   d)  && $95\%\qquad 102\%$
              && $-0.23\pm 0.03$ && $0.1212\pm 0.0018$ && $58.7$ &\cr
   &&       && $94\%\qquad 101\%$
              && $-0.20\pm 0.03$ && $0.1202\pm 0.0021$ && $59.7$ &\cr\tablerule
}}\hfil\bigskip
\centerline{\vbox{\hsize= 380pt \raggedright\noindent\footnotefont
Table~7: As table~5, but to demonstrate insensitivity to large $x$
distributions and the fitting of the small $x$ tail: all calculations at
two loops with $Q_0 = 2\GeV$ and
using a) \DZP, b) \DMP, c) \DZP, but now with
$\lambda_q =\lambda$, and $\lambda_g = \lambda+\Delta\lambda$,
$\Delta\lambda = -0.3, 0.3$, d) using the two alternative fitting
procedures described in the text.
}}
\bigskip
\endinsert

\midinsert\hfil
\vbox{\tabskip=0pt \offinterlineskip
      \def\tablerule{\noalign{\hrule}}
      \halign to 350pt{\strut#&\vrule#\tabskip=1em plus2em
                   &\hfil#\hfil&\vrule#
                   &#\hfil&\vrule#
                   &\hfil#&\vrule#
                   &#\hfil&\vrule#
                   &\hfil#&\vrule#\tabskip=0pt\cr\tablerule
      &&\omit&&\omit\hidewidth norm.\hidewidth
             &&\omit\hidewidth $\lambda$\hidewidth
             &&\omit\hidewidth $\alpha_s(M_Z)$\hidewidth
             &&\omit\hidewidth $\chi^2$\hidewidth&\cr\tablerule
   &&   a)  && $95\%\qquad 102\%$
              && $-0.23\pm 0.03$ && $0.1215\pm 0.0017$ && $58.4$ &\cr\tablerule
   &&   b)  && $96\%\qquad 103\%$
              && $-0.25\pm 0.03$ && $0.1215\pm 0.0017$ && $58.2$ &\cr\tablerule
   &&       && $94\%\qquad 101\%$
              && $-0.42\pm 0.03$ && $0.1222\pm 0.0014$ && $71.7$ &\cr
   &&   c)  && $96\%\qquad 103\%$
              && $-0.13\pm 0.04$ && $0.1216\pm 0.0022$ && $58.7$ &\cr
   &&       && $98\%\qquad 104\%$
              && $-0.08\pm 0.05$ && $0.1219\pm 0.0024$ && $60.9$ &\cr\tablerule
   &&       && $95\%\qquad 102\%$
              && $-0.14\pm 0.05$ && $0.1178\pm 0.0024$ && $60.8$ &\cr
   &&   d)  && $95\%\qquad 102\%$
              && $-0.20\pm 0.04$ && $0.1187\pm 0.0024$ && $58.7$ &\cr
   &&       && $95\%\qquad 102\%$
              && $-0.25\pm 0.03$ && $0.1252\pm 0.0018$ && $58.7$ &\cr
   &&       && $95\%\qquad 102\%$
              && $-0.27\pm 0.03$ && $0.1295\pm 0.0024$ && $59.5$ &\cr
\tablerule
}}\hfil\bigskip
\centerline{\vbox{\hsize= 380pt \raggedright\noindent\footnotefont
Table~8: As table~5, but to show the scheme dependence: all calculations
at two loops with $Q_0 = 2\GeV$ and using a) \MS\ evolution,
b) DIS evolution, c) as a), but with factorization scale $k_1 = \half,2,4$,
d) as a) but with renormalization scale $k_2 = \quarter,\half,2,4$.
}}
\bigskip
\endinsert

\midinsert\hfil
\vbox{\tabskip=0pt \offinterlineskip
      \def\tablerule{\noalign{\hrule}}
      \halign to 350pt{\strut#&\vrule#\tabskip=1em plus2em
                   &\hfil#\hfil&\vrule#
                   &#\hfil&\vrule#
                   &\hfil#&\vrule#
                   &#\hfil&\vrule#
                   &\hfil#&\vrule#\tabskip=0pt\cr\tablerule
             &&\omit\hidewidth $Q_{\rm cut}$\hidewidth
             &&\omit\hidewidth norm.\hidewidth
             &&\omit\hidewidth $\lambda$\hidewidth
             &&\omit\hidewidth $\alpha_s(M_Z)$\hidewidth
             &&\omit\hidewidth $\chi^2$\hidewidth&\cr\tablerule
   &&   10  && $95\%\qquad 102\%$
              && $-0.24\pm 0.04$ && $0.1213\pm 0.0015$ && $45/92$ &\cr
   &&   5  && $94\%\qquad 102\%$
              && $-0.23\pm 0.04$ && $0.1211\pm 0.0024$ && $22/53$ &\cr
   &&   4  && $96\%\qquad 101\%$
              && $-0.22\pm 0.06$ && $0.1225\pm 0.0031$ && $11/33$ &\cr
   &&   3.5  && $98\%\qquad 100\%$
              && $-0.24\pm 0.07$ && $0.1220\pm 0.0043$ && $6/20$ &\cr
   &&   3  && $101\%\qquad 99\%$
              && $-0.33\pm 0.13$ && $0.1142\pm 0.0177$ && $3/7$ &\cr
   \tablerule
   &&   3  && $95\%\qquad 103\%$
              && $-0.22\pm 0.05$ && $0.1219\pm 0.0023$ && $54/111$ &\cr
   &&   4   && $96\%\qquad 103\%$
              && $-0.21\pm 0.10$ && $0.1221\pm 0.0034$ && $48/85$ &\cr
   &&   5  && $96\%\qquad 103\%$
              && $-0.27\pm 0.21$ && $0.1216\pm 0.0039$ && $44/77$ &\cr
   &&   7  && $99\%\qquad 101\%$
              && $-0.39\pm 0.35$ && $0.1211\pm 0.0064$ && $26/53$ &\cr
\tablerule
}}\hfil\bigskip
\centerline{\vbox{\hsize= 380pt \raggedright\noindent\footnotefont
Table~9: As table~5, but to show the stability as the data are cut in
$Q$: in the upper half of the table data above $Q_{\rm cut}$ are
removed, while in the lower half data below $Q_{\rm cut}$ are
removed.
}}
\bigskip
\endinsert

\midinsert\hfil
\vbox{\tabskip=0pt \offinterlineskip
      \def\tablerule{\noalign{\hrule}}
      \halign to 280pt{\strut#&\vrule#\tabskip=1em plus2em
                  &\hfil#\hfil&\vrule#
                   &\hfil#&\vrule#\tabskip=0pt\cr\tablerule
   && two-parameter fit             && $\pm{0.003}$  &\cr
   && variation of $Q_0$            && $\pm{0.002}$  &\cr
   && fitting to large-$x$ p.d.f.   && $\pm{0.003}$  &\cr\tablerule
   && higher order singularities   && $\epm{0.002}{0.006}$  &\cr
   && position of thresholds   && $\pm{0.002}$  &\cr
   && form of running coupling   && $\pm{0.001}$  &\cr
   && higher twist corrections   && $\pm{0.001}$  &\cr
   && scheme dependence   && $\epm{0.008}{0.004}$  &\cr
\tablerule
}}\hfil\bigskip
\centerline{\vbox{\hsize= 380pt \raggedright\noindent\footnotefont
Table~10: Errors in the determination of $\alpha_s(M_Z)$.
}}
\bigskip
\endinsert

\vfill
\supereject
\nopagenumbers
\vsize=27truecm
\null
\vskip -4.truecm
\epsfxsize=14truecm
\hfil\epsfbox{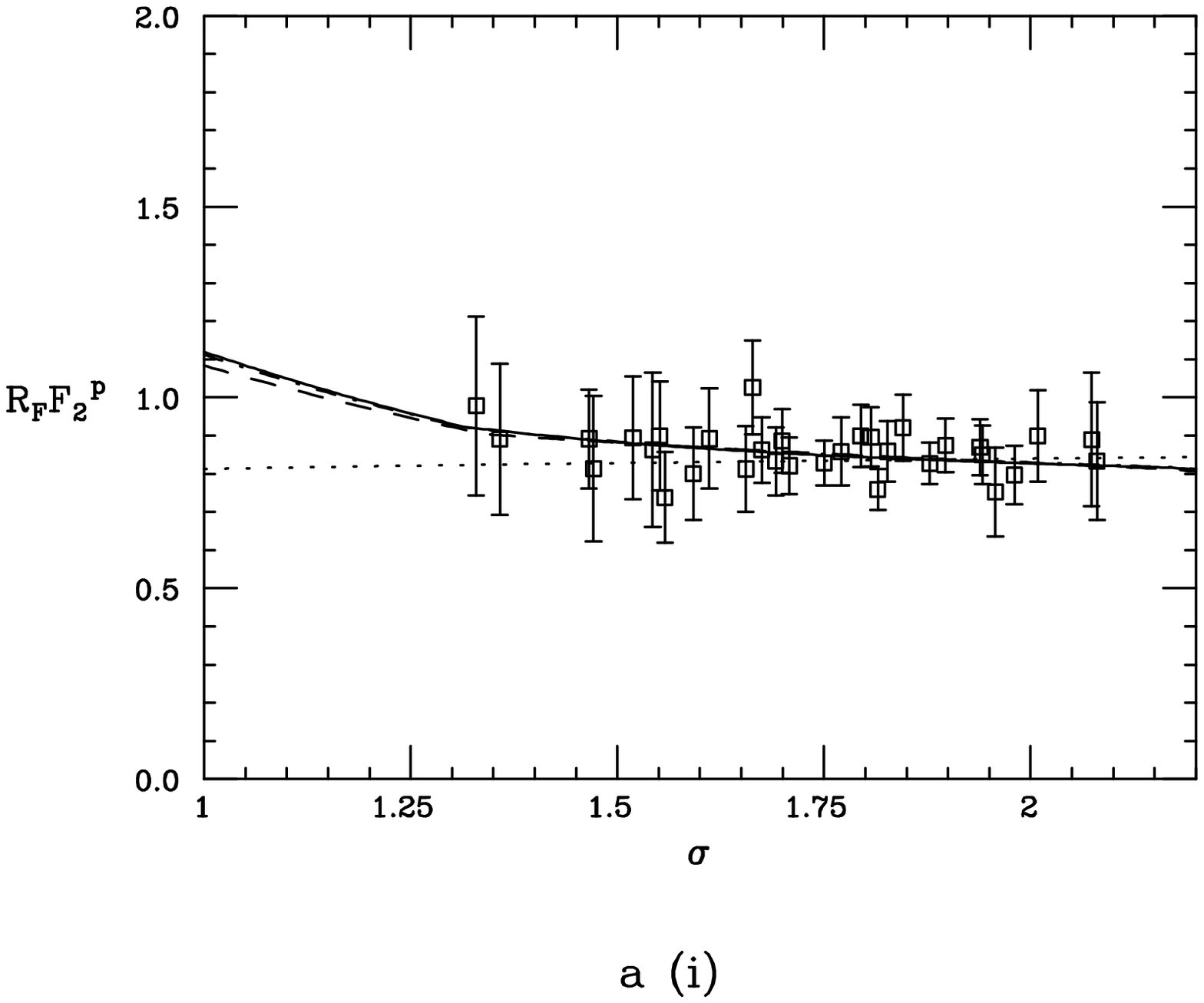}\hfil
\smallskip
\vskip -7.truecm
\epsfxsize=14truecm
\hfil\epsfbox{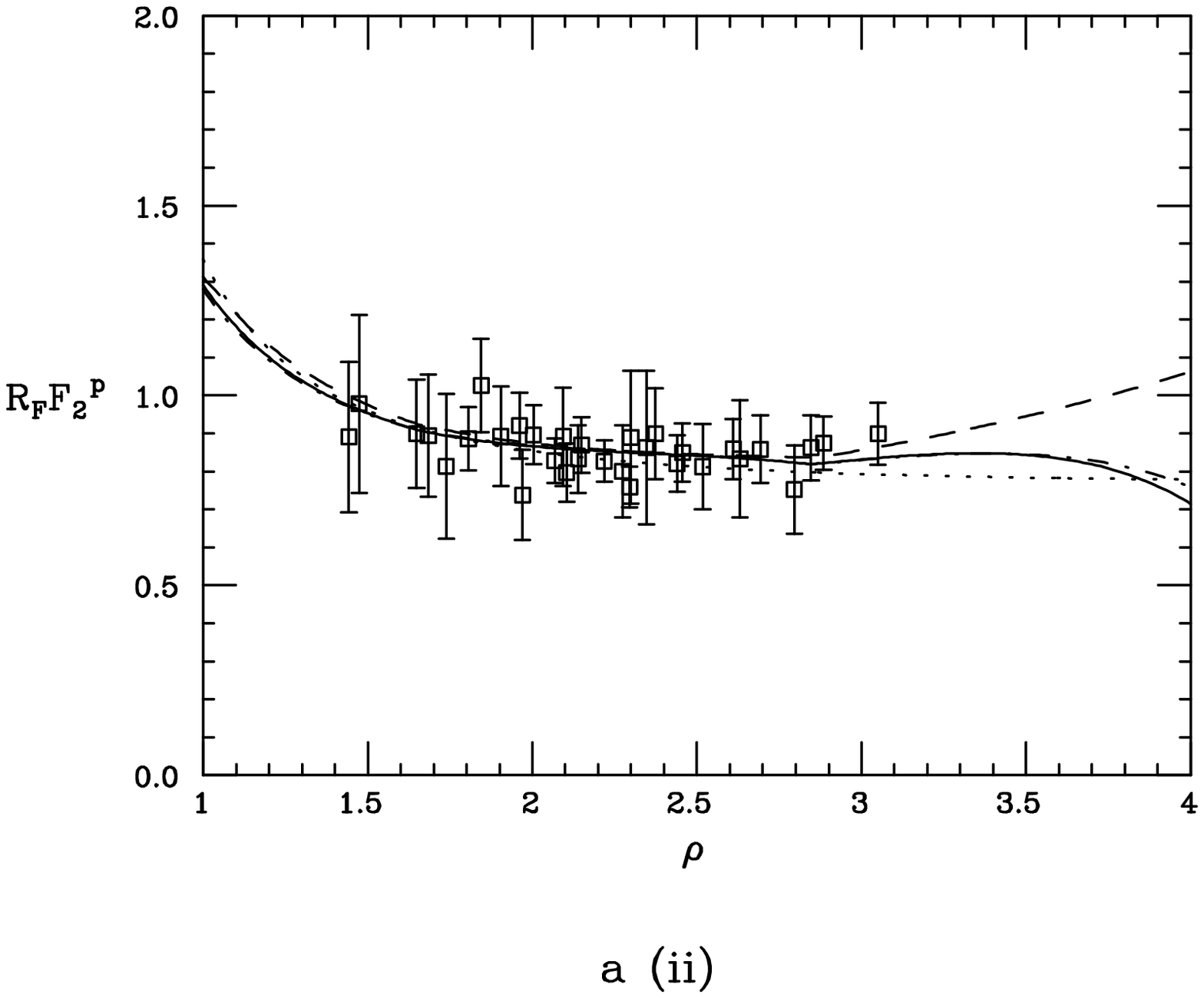}\hfil
\vskip -3.truecm
\centerline{\qquad Fig.~1}
\vfill
\eject
\null
\vskip -4.truecm
\epsfxsize=14truecm
\hfil\epsfbox{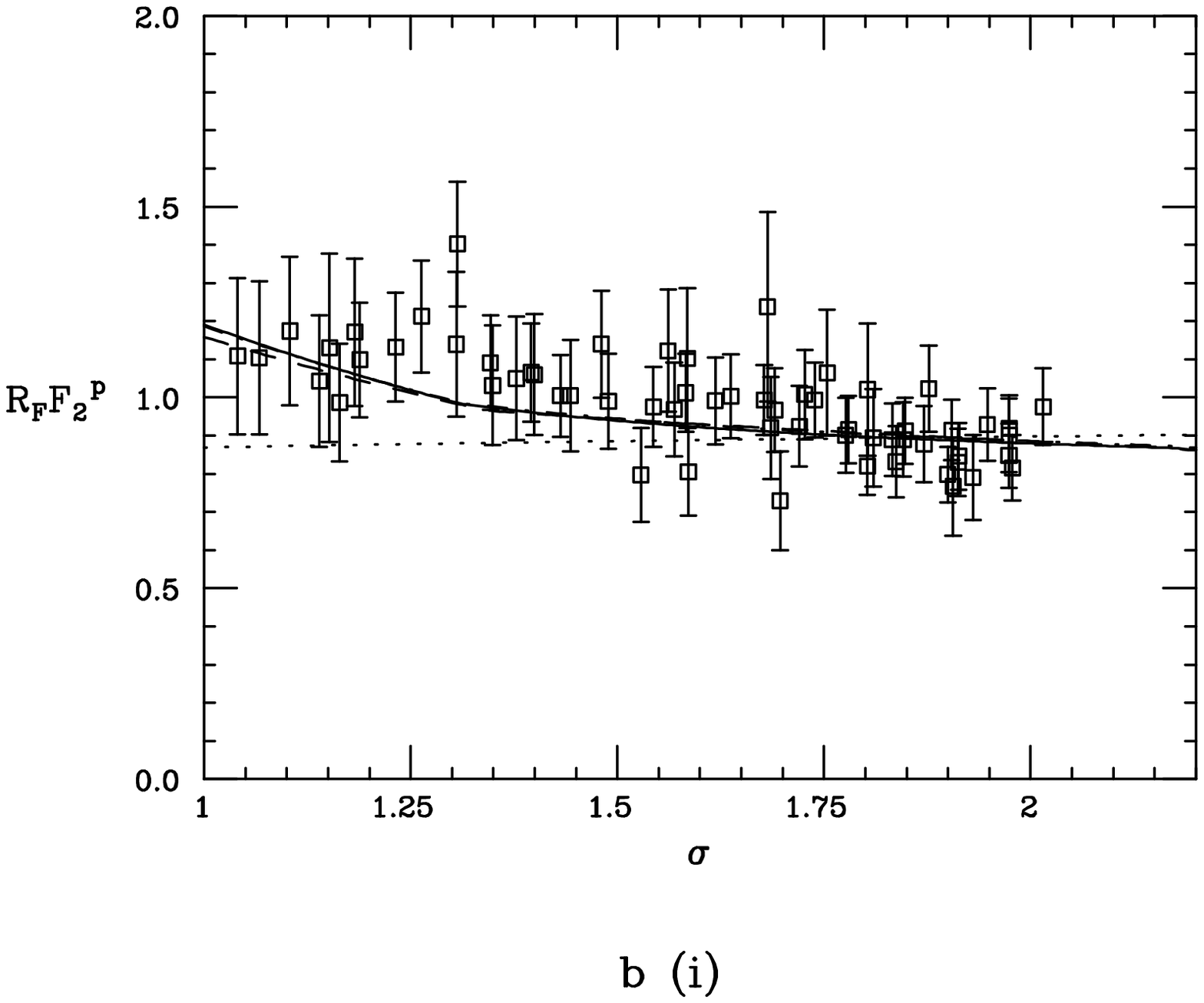}\hfil
\smallskip
\vskip -7.truecm
\epsfxsize=14truecm
\hfil\epsfbox{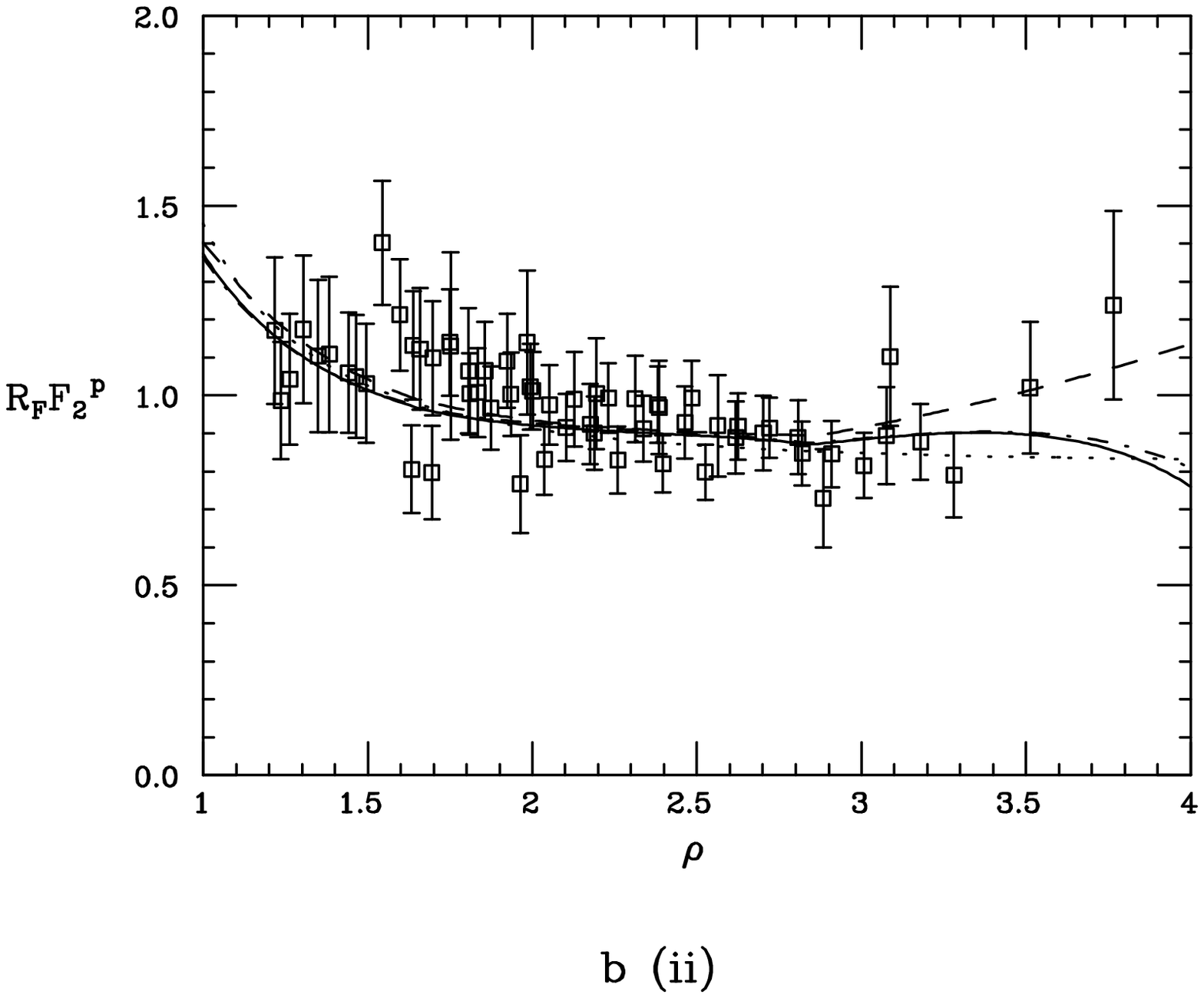}\hfil
\vskip -3.truecm
\centerline{\qquad Fig.~1}
\vfill
\eject
\null
\epsfxsize=8.truecm
\hfil\epsfbox{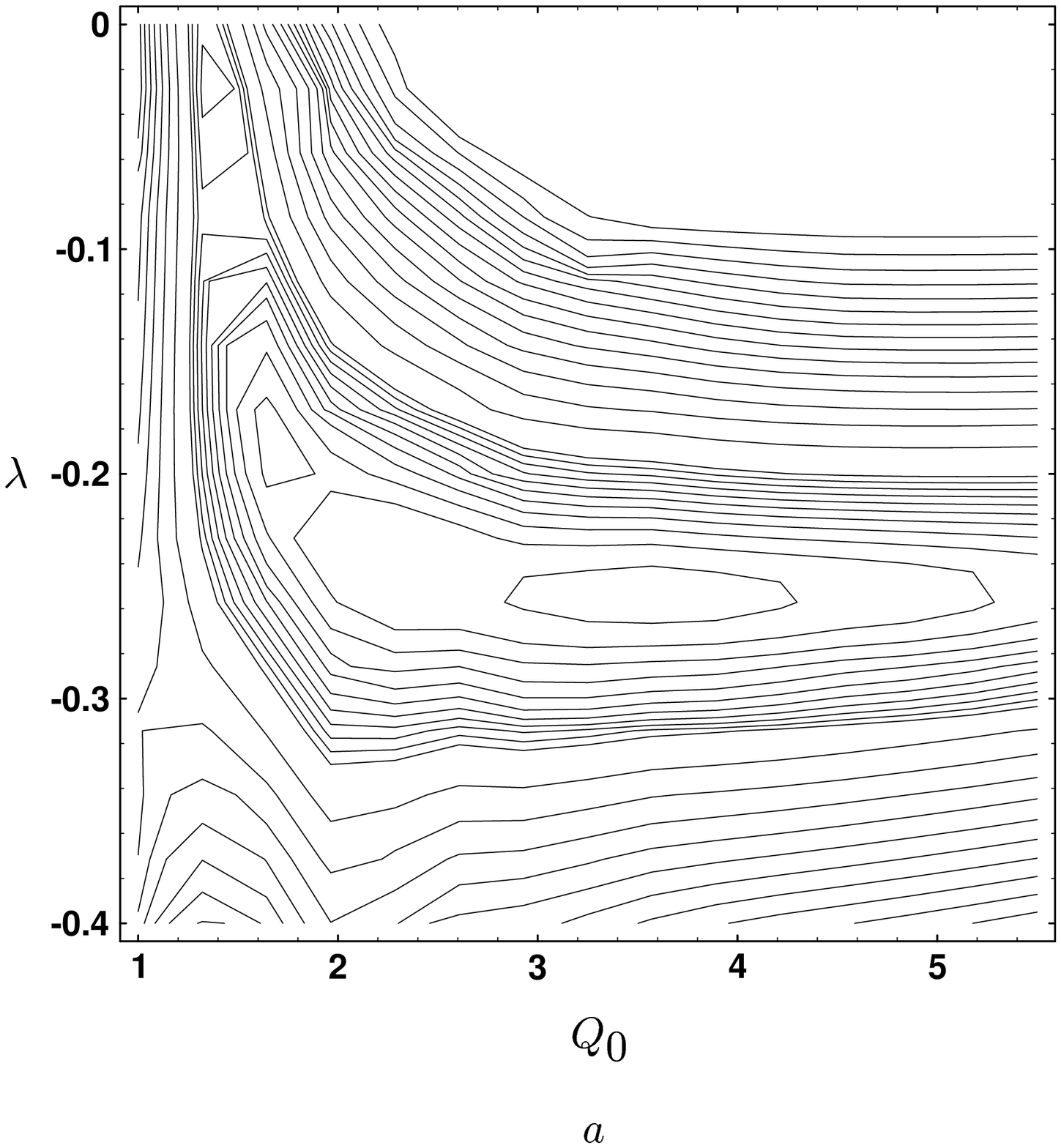}\hfil
\epsfxsize=8truecm
\hfil\epsfbox{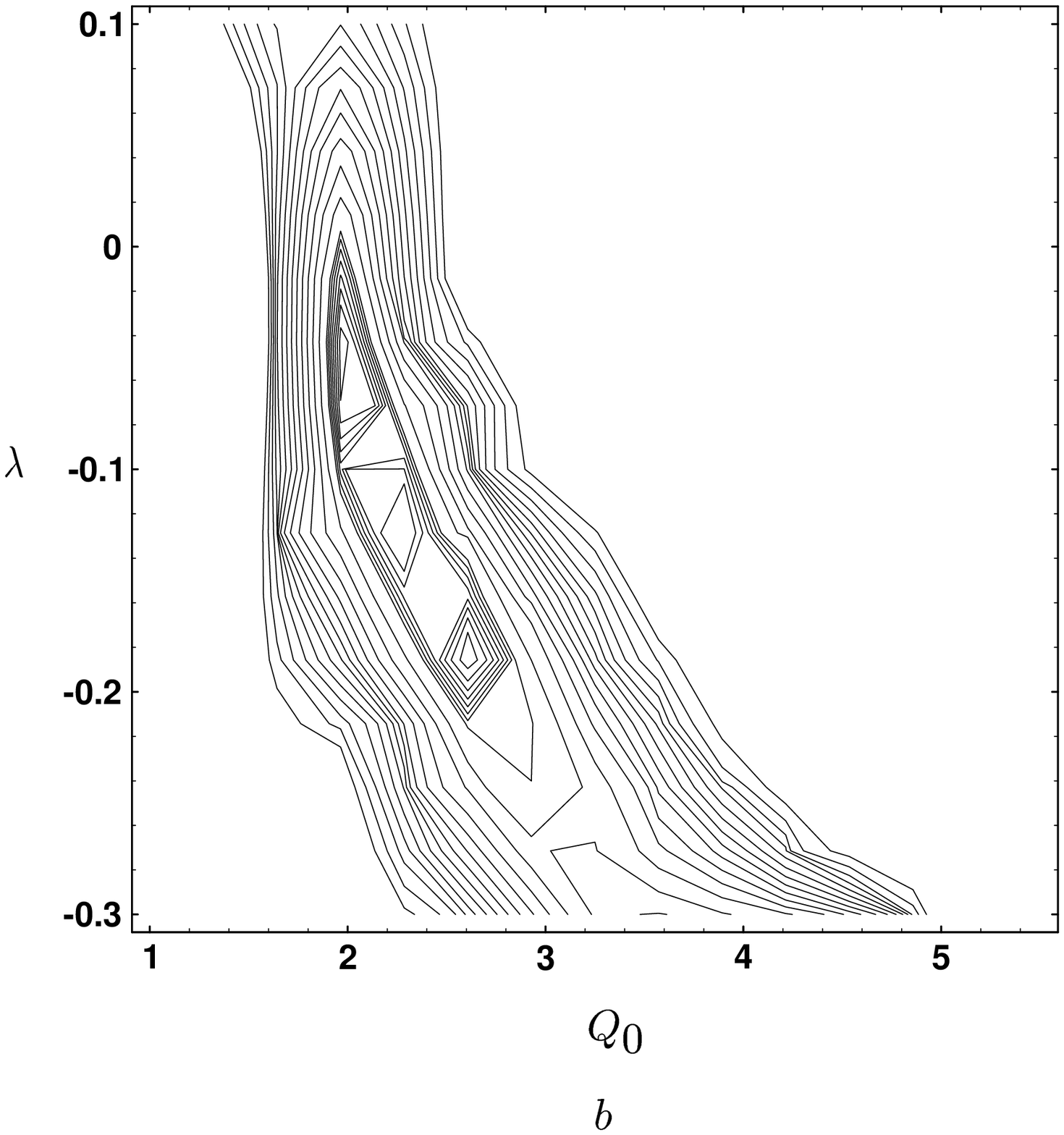}\hfil
\centerline{\qquad Fig.~2}
\epsfxsize=8truecm
\hfil\epsfbox{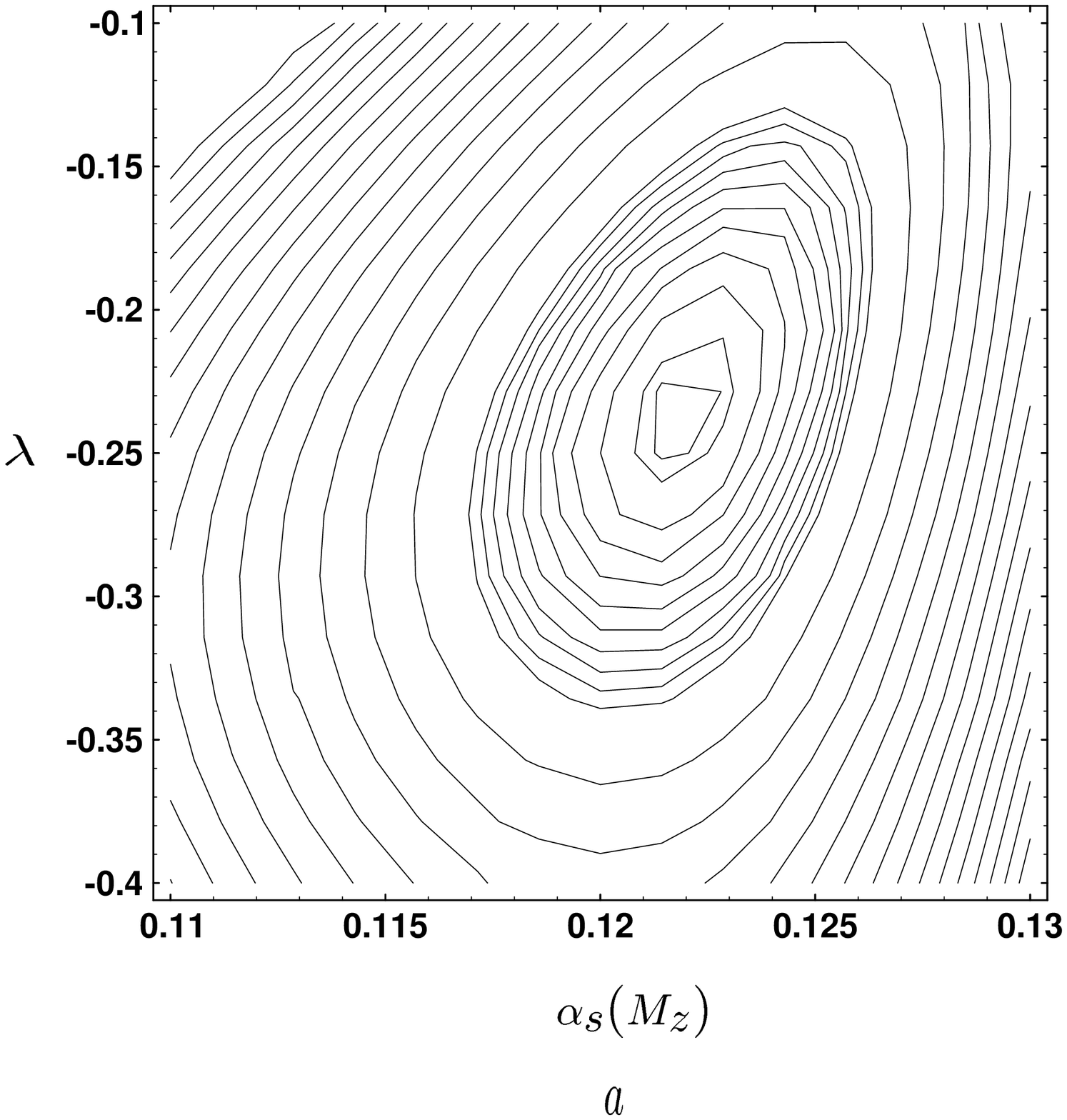}\hfil
\epsfxsize=8truecm
\hfil\epsfbox{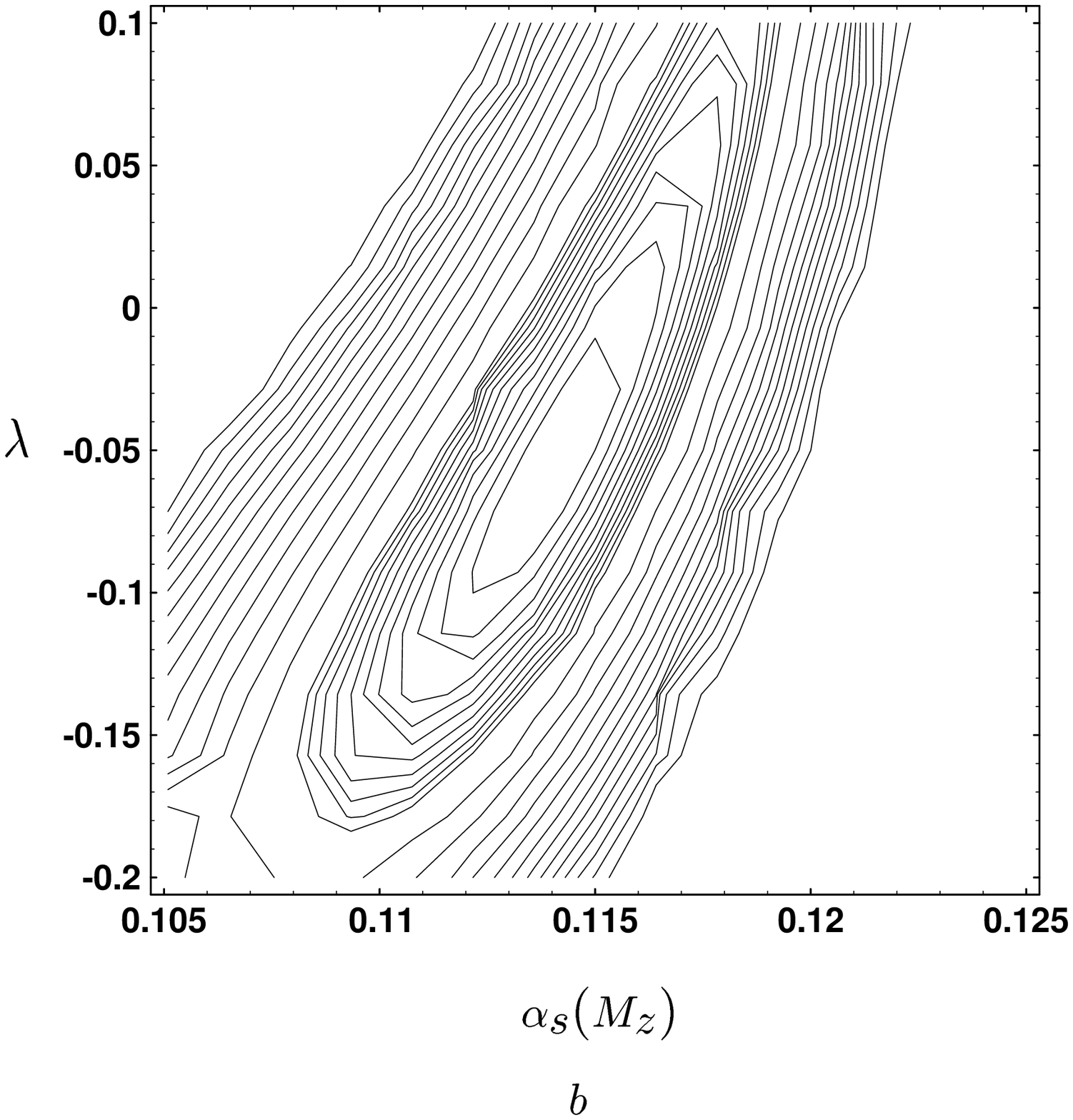}\hfil
\centerline{\qquad Fig.~3}
\vfill
\eject
\epsfxsize=8truecm
\hfil\epsfbox{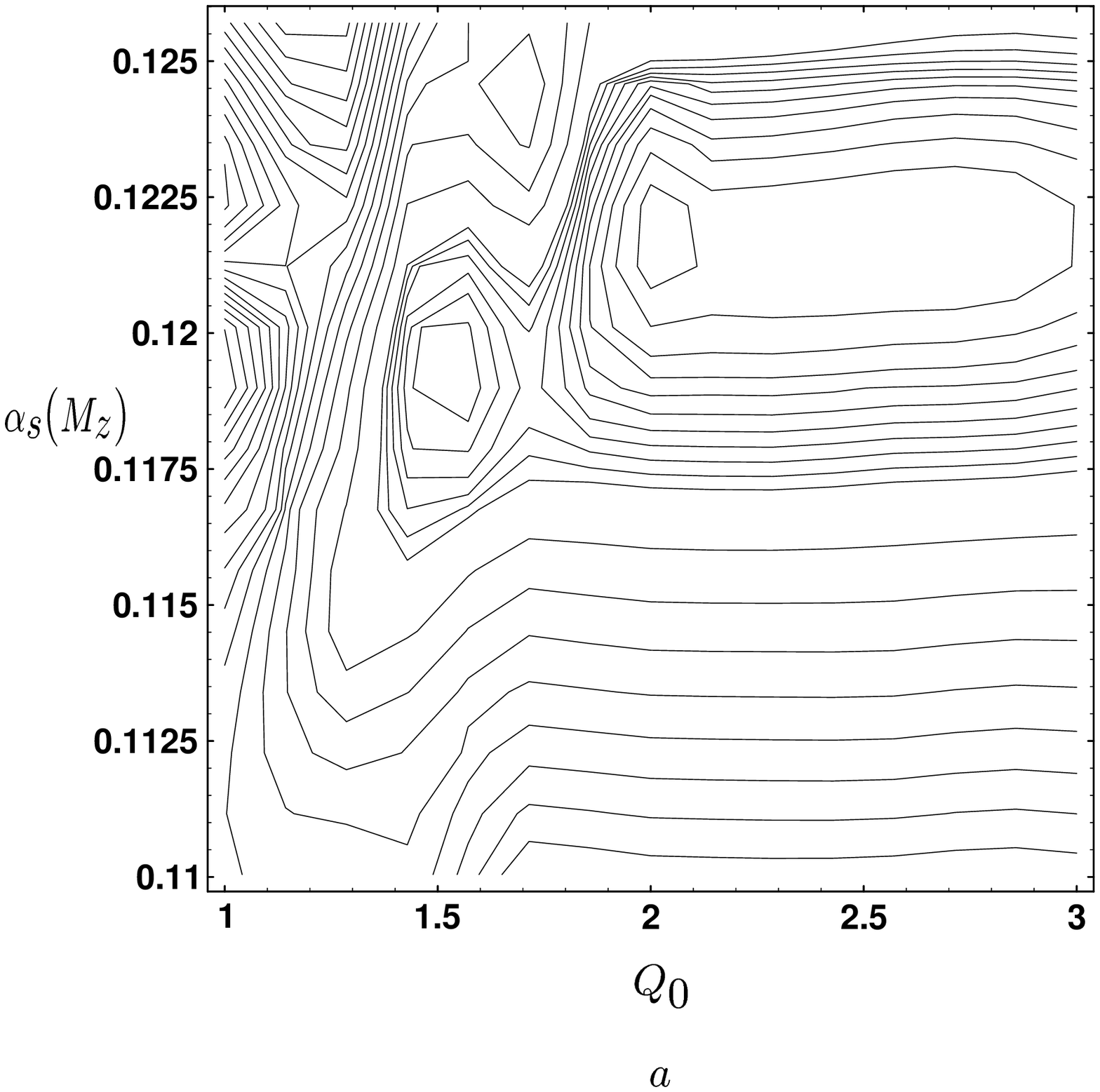}\hfil
\epsfxsize=8truecm
\hfil\epsfbox{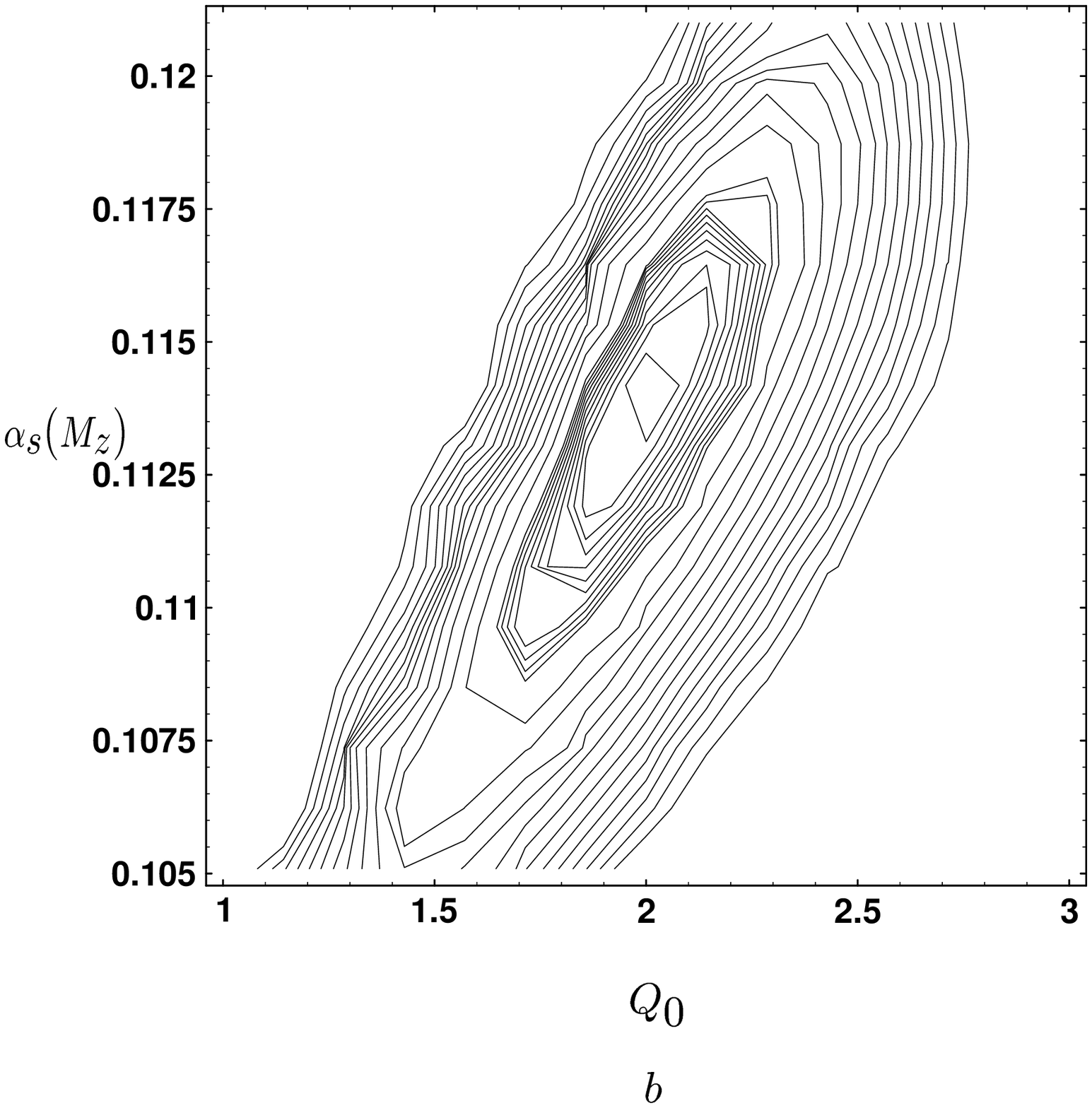}\hfil
\centerline{\qquad Fig.~4}
\vfill
\eject
\bye